\documentclass[journal]{new-aiaa} 
\usepackage{comment}

\usepackage{graphicx}
\usepackage{amsmath}
\usepackage[version=4]{mhchem}
\usepackage{siunitx}
\usepackage{longtable,tabularx}
\usepackage{amsmath, amssymb}
\usepackage{bbm}
\usepackage[english]{babel}
\usepackage{algorithm}
\usepackage{algorithmic}

\newcommand{\STAB}[1]{\begin{tabular}{@{}c@{}}#1\end{tabular}}
\newcommand{\T}{{\scriptscriptstyle\mathsf{T}}}

\newcommand{\argmax}{\mathop{\mathrm{argmax}}}

\newcommand{\nor}[1]{\left\| #1 \right\|} 
\newcommand{\snor}[1]{\left| #1 \right|} 
\newcommand{\LRp}[1]{\!\left( #1 \right)} 
\newcommand{\LRs}[1]{\!\left[ #1 \right]} 
\newcommand{\LRc}[1]{\!\left\{ #1 \right\}} 

\newcommand{\mc}[1]{\mathcal{#1}} 
\newcommand{\mb}[1]{\mathbf{#1}} 
\newcommand{\mbb}[1]{\mathbb{#1}} 

\newcommand{\bs}[1]{\boldsymbol{#1}}

\title{Dimension Bridging for 3D RANS with Neural Network Accelerated Gaussian Functional Regression}

\author{Wesley Lao \footnote{Ph.D. Student, Oden Institute for Computational Engineering \& Sciences, 201 E 24th Street, Austin, TX 78712} \footnote{Corresponding author, email: \href{emailto:wesley.lao@austin.utexas.edu}{wesley.lao@austin.utexas.edu}}, Thomas Scott \footnote{Ph.D. Student, Department of Aerospace Engineering \& Engineering Mechanics, 2617 Wichita Street, C0600
Austin, Texas 78712.}, and Tan Bui \footnote{Professor, Department of Aerospace Engineering \& Engineering Mechanics and Oden Institute for Computational Engineering \& Sciences, 201 E 24th Street, Austin, TX 78712.}}
\affil{University of Texas at Austin, Austin, TX 78712}
\author{Paul Laiu \footnote{Staff Mathematician, Oak Ridge National Laboratory, 1 Bethel Valley Road,
Oak Ridge, TN 37830.} and Matt Bement \footnote{Section Head, Oak Ridge National Laboratory, 1 Bethel Valley Road,
Oak Ridge, TN 37830.}}
\affil{Oak Ridge National Laboratory, Oak Ridge, TN 37830}

\begin{document}

\maketitle

\begin{abstract}
    In many computational science and engineering problems, repeatedly solving fully resolved physics-based models to design for a quantity of interest (QoI) can quickly become intractable, requiring the use of low-fidelity
    models to predict the same QoI but introduce errors where some features are neglected or are otherwise inaccurately resolved.
    We use Gaussian Functional Regression (GFR) to learn a correction to a 2D Reynolds-Averaged Navier-Stokes (RANS) model to predict the aerodynamic coefficients from a 3D RANS model.
    This model pair has a disparity in the governing physics from the reduced dimensionality, a previously unexplored application for GFR.
    Empirically, our results show that with a proper choice of low-dimensional (LD) model, the proposed kernel allows for the use of fewer high-dimensional (HD) evaluations to regress a response surface to the same level of accuracy as standard stationary kernels. Moreover, the new kernel provides more informative uncertainty quantification, which we show is advantageous when used to drive an adaptive sampling algorithm. Finally, we propose a novel neural network accelerated kernel, which we show offers predictions in good agreement while speeding up evaluations by millions of times in wall clock measurements, bringing the computational budget within the real-time regime.
\end{abstract} 

\section{Introduction}
In many design and decision-making applications, a predictive model is iteratively evaluated to find design parameters
which optimize a performance indicator or other QoI. This introduces a tradeoff between the fidelity of the
predictive model and the computational cost to perform the optimization procedure. 
It is common to combine predictions from a set of models of varying fidelity (and cost) such that the design loop has a lower total
computational cost than if only the high-fidelity model were used \cite{fernandez2016review,peherstorfersurvey2018}.
High-fidelity models refer to finely discretized solutions to the PDEs
which govern a physical application. Common low-fidelity models include coarse discretizations, simplified physics
\cite{nguyen2015gaussian,nguyenfunctional2016,nguyengaussian2016,morrison2018representing},
lower dimensionality \cite{morrison2018representing}, projected basis reduced-order models \cite{rozza2008reduced,
nguyengaussian2016}, and data-driven surrogates \cite{jones1998efficient,nguyen2015gaussian,nguyenfunctional2016,
nguyengaussian2016}.

We are interested in predicting and optimizing the performance of a parametric design geometry where the associated
high-fidelity model is prohibited due to its cost for such a many-query task. This places a limit on the availability of high-fidelity
data we can use to regress QoI values as a function of the design and operation parameters (also called a response
surface \cite{jones1998efficient}), which prevents us from directly fitting data-driven surrogates to high-fidelity
data. Instead, we use the predictions of a low-fidelity model, and introduce corrections using the high-fidelity data to better capture the underlying physics.
We also desire uncertainty quantification, for which we turn to Gaussian Process Regression (GPR)
\cite{williams2006gaussian}, also known as \emph{kriging} in geostatistics \cite{matheron1963principles},
which provides a response surface prediction and uncertainty quantification as mean and pointwise standard deviation,
respectively, of a Gaussian Bayesian posterior distribution over the input space.

In particular, we assume that our low-fidelity model is cheap enough to be run at every point in our design space.
In the Bayesian sense, the low-fidelity response surface will serve as the mean of our prior distribution. Similar to
\cite{kennedy2000predicting}, we will add a Gaussian process (GP) correction such that the posterior mean approximates the
high-fidelity response surface. However, rather than standard stationary kernels typically used in GPR, we instead
desire to use the low-fidelity physics to better inform the regression. To this end, we consider the addition of
a GP forcing term in the low-fidelity PDE. In the weak form, this forcing term is identified by a linear
functional, thus allowing us to use the machinery of Gaussian Functional Regression (GFR) developed in
\cite{nguyen2015gaussian, nguyenfunctional2016,nguyengaussian2016} to condition the forcing term on evaluations of
various observation functionals. This is equivalent to using the low-fidelity model as a feature map, with which the
response surface kernel is composed (see Appendix \ref{app:kernel} for a formal derivation.). This featurization allows
the kernel to identify where the low-fidelity response has higher sensitivity (which we assume is related to uncertainty) and gives a better set of basis functions which, when weighted by the observed data, regress the response surface.

In the zero-noise case, we would consider this basis to be a ``cardinal basis'', in which each basis function is equal to 1 at the point corresponding to its observation weight and 0 at other observation points. It is easy to see that this would be an interpolation of the observations. With noise, our basis may not exactly recover this cardinality property (i.e., it is a regression of the observations). Equivalently, one may consider the interpretation that the kernel provides a better basis of kernel functions to regress the response. However, the relationship between the kernel function basis and data is not immediately obvious in the latter form, since the inverse covariance maps the data to the weights on the kernel basis, as shown later in Eq. \eqref{eq:gp_post_mean}. Herein, our visualizations will be using the data-weighted basis interpretation.

In this work, we will expand the use of GFR to nonlinear low-fidelity models and nonlinear observation operators which lie in a lower-dimensional physical
space. This construction places the majority of the compute on the low-dimensional model, so we describe the models by their physical dimensionality as high-dimensional (HD) or low-dimensional (LD), as opposed to their relative fidelity. Since our goal is to connect the cost of an LD model with the accuracy of an HD one, we call this approach a \emph{dimension-bridging} model.

In the case where the LD model or geometry is not derived from symmetry of the HD model, such
that changes in a subset of design parameters cannot be captured by the (primary) LD model, we will introduce an
auxiliary LD model that can capture the neglected parameters. The combination of primary and auxiliary
LD models will give rise to a separable featurized kernel, which will be used to regress the HD
response surface. Further, we will show that driving an adaptive sampling algorithm with the uncertainty quantification
given by the featurized kernel leads to faster convergence of the regressed surface relative to standard stationary
kernels.

Even with use of a significantly cheaper LD model, evaluations of this kernel can be prohibitively expensive.
Looking ahead to our test problem, we will require a 2D Compressible Reynolds-Averaged Navier Stokes (RANS) forward and adjoint solve for every new parameter.

The need for fast inference of the QoIs in applications such as simulating flight dynamics using the force and moment coefficients motivates our proposed novel neural network (NN) approach for learning an arbitrary Mercer kernel. 
The NN training is comparable in cost to the GF hyperparameter tuning, thus not increasing offline cost significantly.
More importantly, during online inference when evaluations of unseen parameters are required, the neural network evaluation will remove the need for additional RANS solves in the online phase. The speedup in terms of wall clock evaluation time is on the order of millions.
Moreover, these evaluations perform well both in accuracy and speed when forming and factoring kernel matrices for predicting the GP posterior.
Ultimately our integration of the physics-informed GF correction with a high-speed NN Mercer kernel surrogate enables LD-physics-informed real-time predictions of the HD-QoIs.

The rest of the paper is organized as follows: Section \ref{sec:background} provides background knowledge on GPR and GFR, Section \ref{sec:problem} gives a
mathematical statement of the regression problem, Section \ref{sec:kernel} describes the construction of the proposed featurized
kernel and hyperparameter selection, Section \ref{sec:adapt} covers our adaptive sampling strategy, Section \ref{sec:nn} details the neural network for accelerating evaluations, Section \ref{sec:algorithm} presents the algorithms, Section \ref{sec:results}
compares our proposed kernel construction and standard GPR kernels when applied
to response surface regression and adaptive sampling, and concluding remarks are presented in Section \ref{sec:conclusion}.

\section{Background and Relevant Work}\label{sec:background}
\subsection{Gaussian Process Regression}
We begin by briefly reviewing GPR (kriging). GPR is a statistical data-driven surrogate model, which seeks to approximate a
function by considering it to be a realization of a GP. A GP is a distribution over functions and can
be thought of as a generalization of the finite multivariate Gaussian distribution. Analogous to multivariate Gaussians, GPs
are uniquely characterized by a mean and covariance. In this case, however, the mean is a measurable function and the
covariance is a symmetric, positive-definite kernel.

GPR involves constructing (generating) a joint distribution for the evaluation of the function at training and testing points.
We assign a GP prior on $f$, with some mean function $m$ and covariance kernel $k$. The joint distribution is then generated
via pointwise evaluation of $m$ and $k$, i.e.,

\begin{equation}
    \begin{bmatrix}
        f\LRp{\bs{x}} \\
        f\LRp{\hat{\bs{x}}}
    \end{bmatrix} \sim \mathcal{N}\LRp{\begin{bmatrix}
        m\LRp{\bs{x}} \\
        m\LRp{\hat{\bs{x}}}
    \end{bmatrix},\,\begin{bmatrix}
        k\LRp{\bs{x},\bs{x}} & k\LRp{\bs{x},\hat{\bs{x}}} \\
        k\LRp{\hat{\bs{x}},\bs{x}} & k\LRp{\hat{\bs{x}},\hat{\bs{x}}} \\
    \end{bmatrix}}
\end{equation}

where $\bs{x}$ is the vector of training points and $\hat{\bs{x}}$ is the vector of test points. We abuse the
notation of a function with scalar arguments acting on vector-valued inputs to denote the generation of a vector
by evaluating the function on the argument(s) element-wise. Evaluating the kernel on elements of the input vectors in lexicographical order generates a matrix. We observe the (noisy) evaluation of $f\LRp{\bs{x}} +
\bs{\eta} =: \bs{d}$, where noise vector $\bs{\eta}$ is assumed to be i.i.d. from a centered Gaussian with
standard deviation $\sigma$. Then, predictions on test points amounts to computing the conditional distribution of the
evaluation of $f\LRp{\hat{\bs{x}}}$, given the observations $\bs{d}$. Numerically, this is nothing but a
Schur complement, where the covariance matrix of the training evaluations need only be inverted once before predicting on any number of test
point sets.

\begin{subequations}
    \begin{gather}
        \LRp{f\LRp{\hat{\bs{x}}}|f\LRp{\bs{x}}=\bs{d} - \bs{\eta}} \sim \mathcal{N}\LRp{m^\dagger\LRp{\hat{\bs{x}}},\,k^\dagger\LRp{\hat{\bs{x}},\hat{\bs{x}}}} \\
        m^\dagger\LRp{\hat{\bs{x}}} = m\LRp{\hat{\bs{x}}} + k\LRp{\hat{\bs{x}},\bs{x}}K_d^{-1}(\bs{d} - m\LRp{\bs{x}})
        \label{eq:gp_post_mean}\\
        \label{eq:gp_post_cov}
        k^\dagger\LRp{\hat{\bs{x}},\hat{\bs{x}}} = k\LRp{\hat{\bs{x}},\hat{\bs{x}}} - k\LRp{\hat{\bs{x}},\bs{x}}K_d^{-1}k\LRp{\bs{x},\hat{\bs{x}}} \\
        K_d = \LRp{k\LRp{\bs{x},\bs{x}} + \sigma^2 I}
    \end{gather}
\end{subequations}

Here, our use of the $\dagger$ superscript denotes a posterior quantity. $K_d$ is the covariance matrix of $f\LRp{\bs{x}}+\bs{\eta}$.

\subsection{Gaussian Functionals}
Gaussian Functionals (GFs)  are an extension of GPs to functionals (maps from functions to scalars) \cite{nguyen2015gaussian,
nguyenfunctional2016,nguyengaussian2016}. In the same way that GPs map vectors in $\mbb{R}^N$ to multivariate Gaussians, GFs map vectors of measurable functions in $\mathcal{M}^N$ to multivariate Gaussians. 

The action of a GF $g$ with mean functional $m$ and covariance kernel $k$ on a vector of $N$ test functions $\bs{v} := \LRc{v_i}_{i=1}^N$ is jointly Gaussian:

\begin{equation}
    g\LRp{\bs{v}} = \begin{bmatrix}
            g\LRp{v_1} \\
            \vdots \\
            g\LRp{v_N}
        \end{bmatrix} \sim \mathcal{N}\LRp{m\LRp{\bs{v}},k\LRp{\bs{v},\bs{v}}}
    \label{eq:gfs}
\end{equation}

In this paper, we are interested in regressing a map from design parameters $\bs{\theta}$ to forcing correction $g$ in our low-fidelity model. In a weak setting,
the forcing function is a linear functional in the dual of the test space. It is thus natural to model the class of all forcing functions as a GF that acts on a product space of measurable functions $\mathcal{M}$ and the parameter space $\Theta \subset R^N$.
Specifically, we postulate a GF with mean $m$ and covariance $k$ on the forcing $g$:

\begin{subequations}\label{eq:parametrizedGF}
    \begin{gather}
        g \sim \mathcal{GF}\LRp{m,k},\quad \theta\in \Theta, \quad v \in V \\
        \bs{q} := g\LRp{\bs{\theta};\bs{v}} \sim \mathcal{N}\LRp{m\LRp{\bs{v}, \bs{\theta}},k\LRp{\LRp{\bs{v},\bs{\theta}},\LRp{\bs{v},\bs{\theta}}}}
    \end{gather}
\end{subequations}

where $\bs{\theta}$ is the vector of design configurations and $\bs{v}$ is the vector of test functions. $\bs{q}$ is a multivariate random variable output, which will become our response surface surrogate. We interpret the elementwise evaluation over the vector tuple $\LRp{\bs{v},\bs{\theta}}$ as iterating over the flattened tensor product of the two component vectors. For $N$ functions and $N$ parameter configurations, this flattened vector takes the form

\[
\LRp{\bs{v},\bs{\theta}} = \begin{bmatrix}
    \LRp{v_1,\theta_1} \\
    \vdots \\
    \LRp{v_1,\theta_N} \\
    \LRp{v_2,\theta_1} \\
    \vdots \\
    \LRp{v_N,\theta_N} \\
\end{bmatrix}
\]

Though \eqref{eq:parametrizedGF} is defined for general test functions $\bs{v}$, only a specific choice of $\bs{v}$ is needed for the computation of our QoI. In particular, we can show that (a full derivation is provided in Appendix \ref{app:adj_map}) each evaluation of the
QoI response for a given $\bs{\theta}$ is given by the evaluation of the forcing functional $g\LRp{\bs{\theta}}$ on the adjoint solution $\phi\LRp{\bs{\theta}}$ with respect to the QoI for that $\bs{\theta}$.  Thus, $q$ in \eqref{eq:parametrizedGF} for can be rewritten as a GP over the parameter space, where the test function argument to the GF is restricted to $v = \phi\LRp{\bs{\theta}}$

\begin{equation}
        q \sim \mathcal{GP}\LRp{m_q,k_q} \text{ with } 
        m_q\LRp{\bs{\theta}} = m\LRp{\phi\LRp{\bs{\theta}},\bs{\theta}} \text{ and }
        k_q\LRp{\bs{\theta},\bs{\theta}} = k\LRp{\LRp{\phi\LRp{\bs{\theta}},\bs{\theta}},\LRp{\phi\LRp{\bs{\theta}},\bs{\theta}}}
        \label{eq:fgp2gp}
\end{equation}

Note that a realization of the GP response surface is now only a process over the parameter configuration $\theta$, and that the
LD adjoint solution $\phi$ serves as a feature map with which the kernel is composed to give the GP covariance.

\subsection{Sum and Product Kernels}
To avoid misspecification, we parameterize the parameter-only kernel as a non-negatively weighted sum of distinct classes
of kernels, each of which is individually positive definite, i.e.,

\begin{equation}
    k\LRp{\bs{\theta}_i,\bs{\theta}_j} = \sum_{m=1}^{M} \sigma_m k_m\LRp{\bs{\theta}_i,\bs{\theta}_j}
\end{equation}

where $m$ is the kernel model index, $\sigma_m$ are non-negative weights, and the subscript on $\bs{\theta}$ denotes that the arguments
may be elements of some subset of the design space. For example, Mat\'ern kernels \cite{stein1999interpolation}
are parameterized by a positive real number $\nu$, which allows for processes with this kernel to be mean square differentiable up to
$n < \nu$ times \cite{williams2006gaussian}. We optimize over a mixture of Mat\'ern kernels for $\nu \in \LRc{\frac{1}{2},\frac{3}{2},\frac{5}{2},\infty}$ to infer the regularity (or other structure) needed by the kernel from the data. The hyperparameters of the sum mixture include the hyperparameters of each individual kernel and its associated non-negative weight. 

For regressing over a design space in a Cartesian product space $\Theta \subset \mbb{R}^P$, it is natural to consider a product
kernel where each factor only acts on one dimension of the product space,

\begin{equation}
    k\LRp{\bs{\theta}_i,\bs{\theta}_j} = \prod_{p=1}^{P} k_p\LRp{\LRp{\bs{\theta}_i}_p,\LRp{\bs{\theta}_j}_p}
\end{equation}

where $p$ is the element or dimension index. This allows for efficient inversion and matrix-vector product computation via use
of the Kronecker product if the input set is a regular Cartesian grid in $\mathbb{R}^P$. A product spectral mixture kernel was introduced
in \cite{wilson2014fast}, generalizing the work of \cite{wilson2013gaussian} to product input spaces.

The sum kernel structure can be thought of as a logical OR, where two evaluations are correlated if \emph{any} of the summand
kernels predict correlation. By contrast, the product kernel structure can be thought of as a logical AND, where two evaluations
are correlated if \emph{all} of the factor kernels predict correlation. We can combine these concepts to design kernels based on
a number of simple Boolean assumptions. In our case, we use product kernel structure to separate the GF input space into its function space and parameter space factors, ensuring evaluations are correlated only if the parameters are similar AND they have similar sensitivity under the LD model. We then employ a sum kernel within each so that the factor kernel predicts correlation if \emph{any} of its summand kernels do.

\subsection{Interpolated and Sparse Kernels}
Standard GPs suffer heavily from scalability issues. Not only does the computation and storage have $\mathcal{O}\LRp{N^3}$ and
$\mathcal{O}\LRp{N^2}$ complexity, respectively, for $N$ data points, but kernel functions with large support and length scales
(e.g., the popular Gaussian kernel) can lead to ill-conditioned covariance matrices. To overcome this, there have been methods
for improving the scalability of kernels via approximation or sparse construction. Some approximations use a set of $M$ inducing
points, where $M \ll N$, which reduces the computation and storage complexities to $\mathcal{O}\LRp{M^3+M^2 N}$ and
$\mathcal{O}\LRp{MN+M^2}$, respectively \cite{quinonero2005unifying}. A treatment for incomplete grids was proposed in
\cite{wilson2014fast}, where the training set is augmented with dummy observations that do not impact the posterior. The
augmented data covariance is inverted with a preconditioned conjugate gradient method, which allows for the dummy observations
to be ignored. An improvement known as Structured Kernel Interpolation \cite{wilson2015kernel} factors an approximation of the
full data covariance into a matrix  product $k\LRp{\bs{x},\bs{x}} \approx Wk\LRp{\bs{u},\bs{u}}W^{\T}$, where $W$ is the
interpolation matrix and $\bs{u}$ is the vector of inducing points. This can lower the complexities to $\mathcal{O}\LRp{N +
M\mathrm{log}\LRp{M}}$ and $\mathcal{O}\LRp{N+M}$, respectively, allowing for inducing-data point splits with $M \approx N$
\cite{wang2019exact}. In Section \ref{sec:nn}, we introduce our novel ML kernel as a nonlinear interpolation extending our kernel from the training data, which can be thought of as our inducing points, to any arbitrary points.


\subsection{Nonstationarity through Warping}
A kernel is stationary if it can be defined as a function of the distance $d\LRp{\cdot,\cdot}$ between its inputs. That is,

\[
\exists f: k\LRp{x_i,x_j} = f\LRp{d\LRp{x_i,x_j}}
\]

By contrast, a nonstationary kernel does not have translational symmetry. This can be beneficial if the behavior
of the target function changes over its domain (e.g., is more jagged over some regions and more smooth over others), but nonstationarity tends
to have a more complicated parameterization and is harder to optimize.

Warping refers to modifying kernel functions by either pointwise multiplication with a shape function (warping the output) or
by composing the kernel with a continuous transformation (warping the input). For the former, consider a mixture of experts,
where the input space has been divided among possibly infinite ``expert'' GPs, which serve as the predictive model for its subset
\cite{rasmussen2001infinite, meeds2005alternative}. The sum of these expert kernels, multiplied by the warping functions (or 
expert indicators in this case) is thus a nonstationary kernel. For the latter, consider the shape parameter in Gaussian kernels,
where larger shape parameters lead to more decay in the distance term. If this shape parameter varies in space, we are able to
inject information about varying behavior in space, where quickly-varying regions have larger shape parameters and slowly-varying regions have smaller shape parameters, thus creating a nonstationary kernel \cite{gibbs1998bayesian,bozzini2015interpolation}.

In this work, we use a combination of these two warping approaches, where the parameter-to-adjoint map can be thought of as an input warping,
and the product kernel form can be thought of as PDE-informed output warping on a standard kernel. This is similar to the form used in \cite{nguyengaussian2016}, but we employ a different function kernel that directly follows from the GF correction lying in the dual of the test space, as we will show in Section~\ref{sec:kernel} and Appendix~\ref{app:kernel}.

\begin{equation}
    k\LRp{\bs{\theta}_i,\bs{\theta}_j} = k_{\Theta}\LRp{\bs{\theta}_i,\bs{\theta}_j}\!k_{V}\LRp{\phi\LRp{\bs{\theta}_i},\phi\LRp{\bs{\theta}_j}}
    \label{eq:fgp_product_kernel}
\end{equation}

\section{Problem Statement}\label{sec:problem}
We are interested in predicting a QoI for a physical system as a function of design variables and physical parameters, such as
geometry or non-dimensional quantities. We have access to an HD ground-truth model, from which the QoI can be
derived as an integrated quantity of the solution. However, solving this model for every parameter configuration is not feasible for limited
computational budget.
With limited data from the HD model, we wish to predict the QoI at new parameter configurations using a
cheaper, but inaccurate, LD model. Recall that for our purposes, the ``dimension'' of a model refers to the spatial
dimension of the problem under consideration. To this end, our objective is
\begin{quoting}
to regress a corrective forcing term to the LD model such that the error between the corrected LD QoI and HD QoI is minimized.    
\end{quoting}

Consider a general weak PDE for the HD and LD models defined over open bounded domains $\Omega$ and boundaries $\Gamma$, implicitly
parameterized by a configuration $\bs{\theta}$ from the design space $\Theta$. From here on, $\bs{\theta}$ is omitted as an argument unless
it is needed for clarity.

\begin{equation}
    \begin{aligned}
        \mathrm{Seek}&\ u_H \in U_H: \\
        &\begin{cases}
            \LRp{\mathcal{P}_H\LRp{u_H},v} = \LRp{f_H,v} \quad \forall v \in V_H\\
            \LRp{\mathcal{B}_H\LRp{u_H},w} = \LRp{h_H,w} \quad \forall w \in W_H\\
        \end{cases}\\
        \mathrm{Seek}&\ u_L \in U_L: \\
        &\begin{cases}
            \LRp{\mathcal{P}_L\LRp{u_L},v} = \LRp{f_L,v} \quad \forall v \in V_L\\
            \LRp{\mathcal{B}_L\LRp{u_L},w} = \LRp{h_L,w} \quad \forall w \in W_L\\
        \end{cases}
    \end{aligned}
    \label{eq:hd_ld_pde}
\end{equation}

where $\LRp{\cdot,\,\cdot}$ denotes the $L^2$ inner product (or more generally a duality pairing), and the subscript denotes which model the
term is attributed to, namely, $H$ is for HD and $L$ is for LD. Here,  $u$ is    the solution to the PDE, $\mathcal{P}$ and $\mathcal{B}$ are the differential operators on the
domain and boundary, $f$ and $h$ are the forcing and boundary terms, $U$ is the trial space, and $V$ and $W$ are the
test spaces for the interior and boundaries, respectively. 

The QoI $q$ is found by integrating the solution

\begin{equation}
        q\LRp{\bs{\theta}} = F\LRp{u\LRp{\bs{\theta}}}
    \label{eq:qoi}
\end{equation}

where $F$ is an operator that defines QoI $q$ as a function of $u$. 
In general, $q_L \neq q_H$, and we are interested in finding a corrected counterpart of $q_L$, denoted as $q^\dagger$, so that $q^\dagger$ approximates the unknown $q_H$. A straightforward approach would be to construct a GP for the error $e = q_H - q_L$, and then define $q^\dagger = q_L\LRp{\bs{\theta}} + e\LRp{\bs{\theta}}$. This approach is purely data-driven and does not take the underlying structure of the problem into account. For that reason, our numerical examples will show that it could take more data to yield a desired accuracy. Aiming at overcoming this potential issue, we developed a physics-aware approach by postulating a corrective forcing term that indirectly corrects $q_L$ by way of producing $u^\dagger$ by accounting for the missing physics that are relevant to the QoI. 
To that end, we seek $g \in V_L^*\LRp{\Omega_L},\,u^\dagger \in U_L\LRp{\Omega_L}$ such that for all $\bs{\theta}$

\begin{subequations}
        \begin{gather}
            \LRp{\mathcal{P}_L\LRp{u^\dagger},v} = \LRp{f_L,v} + \LRp{g,v} \quad \forall v \in V_L\\
            \LRp{\mathcal{B}_L\LRp{u^\dagger},w} = \LRp{h_L,w} \quad \forall w \in W_L\\
            q^\dagger = F\LRp{u^\dagger}
            \label{eq:wish}
    \end{gather}
    \label{eq:corrected_ld_pde}
\end{subequations}

We further assume that the observed data $d\LRp{\bs{\theta}}$ is corrupted by noise $\eta$, i.e., $d\LRp{\bs{\theta}} = q_H\LRp{\bs{\theta}} + \eta$, and that the noise $\eta$ is also the difference between the data $d\LRp{\bs{\theta}}$ and our corrected QoI $q^\dagger\LRp{\bs{\theta}}$:
\begin{equation}
d\LRp{\bs{\theta}} = q^\dagger\LRp{\bs{\theta}} + \eta
\label{eq:noisy_match}    
\end{equation}


We will place a GF prior on $g$, with mean functional $m$ and covariance kernel $k$. This is equivalent to placing a GP prior over the QoI
response surface $q$, with prior mean equal to the LD response surface and covariance featurized by the configuration-to-adjoint
map. We condition the GP on satisfying the observational model \eqref{eq:noisy_match} on the training set, which provides a dimension-bridging surrogate of the true QoI response surface and its uncertainty by way of the posterior mean and covariance, respectively.

\section{The Inner Product Kernel}\label{sec:kernel}
\subsection{Covariance of Inner Products}
From \eqref{eq:gfs}, we know that the covariance of a GF acts symmetrically on two functions and takes a tuple of functions $v,w$  to a real number
in a positive definite manner, quantifying the covariance between the evaluation of the functional on the two arguments.
The covariance ultimately takes the form
of a double integral operator: 

\begin{equation}
    k_V\LRp{v,w}=\iint\!\kappa\LRp{x,y}\ v\LRp{x}\!dx\, w\LRp{y}\!dy.
    \label{eq:gf_cov_2xint}
\end{equation}

Here, the integral kernel $\kappa$ is taken as the covariance kernel of a GP (see Appendix B for a detailed discussion).

\subsection{Auxiliary LD Models}
We now consider the effects of geometry that cannot be captured by the LD model, specifically the case where such geometry can be
described with parameters $\xi$. We can extend our parametric GF to act on a new design space defined as the product of the
LD-visible subspace $\Theta$ and the LD-invisible subspace $\Xi$, i.e.,

\begin{subequations}\label{eq:fgp_unseen}
    \begin{gather}
        g \sim \mathcal{GF}\LRp{m,k}, \\
        m: V \times \Theta \times \Xi \rightarrow \mathbb{R}, \\
        k: \LRp{V \times \Theta \times \Xi}^2 \rightarrow \mathbb{R}.
    \end{gather}    
\end{subequations}

However, since the geometry $\xi\in\Xi$ is neglected by the primary LD model, the LD solution and adjoints will not change as $\xi$ varies. This
places all variation in the predicted quantity due to geometry on the geometry kernel $k_\Xi$. Our goal is to exploit the LD
physics solver available to us to design a geometry kernel with improved performance over standard GPR kernels.

Assuming the parametric geometry also admits an LD representation, we may add an auxiliary LD model to featurize $\xi$. This time, we
assume the effects of the geometry are associated with the structure of the new auxiliary LD solution, as opposed to the adjoints.
This is because we are not evaluating the GF on an adjoint solution, but rather using the auxiliary solve as a way to map the parameters that are invisible to the primary LD model to a more feature-rich space which may better inform the regression of our QoI. With this in mind,
we define a new product kernel with factor covariances $k_U$ and dependence on $\Xi$ acting on adjoint solutions, primal solutions, and parameters,
respectively.

\begin{equation}
    k = k_V k_{\Theta \times \Xi}  k_U
    \label{eq:four_kernels}
\end{equation}

where $u \in U$ are the auxiliary solutions associated with the geometric parameters $\xi$. Since $k_U$ is a kernel over a function
space, we employ the same  integral form used in the primary kernel \eqref{eq:gf_cov_2xint}.
Furthermore, in practice we have found that including the forward solutions from the primal model is beneficial even when an auxiliary model is unnecessary.
In our results, we consider two cases: one varying an LD-visible parameter and one varying an LD-invisible parameter that demands an auxiliary solution.
In the former, we use both $k_V$ and $k_U$ via the primary LD model and in the latter, we only use $k_U$ from the auxiliary model.
Still, the most general case remains where a combination of $k_V$ and $k_U$ may be used given both a primary and an auxiliary LD model.

\subsection{Multiple QoIs}
When modeling multiple quantities (e.g., lift, drag, and pitching moment coefficients), we independently learn different parameterizations of our featurized kernel for each QoI. When $k_V$ is used, only the appropriate adjoint for each QoI is utilized. Although the double integral operator \eqref{eq:gf_cov_2xint} allows for a natural form of cross-covariance by sharing the same integral kernel $\kappa$, we have found in practice that this produces poor results. We observed that stronger sensitivities (larger adjoint state magnitudes) tend to dominate, leading the regression of less sensitive QoIs to ``ignore'' the contribution of observations in favor of predicted covariances with more sensitive QoIs.


\subsection{Hyperparameter Tuning}
Most GP models optimize the hyperparameters through maximum likelihood estimation (MLE). In other words,
inferring an optimal hyperparameter set from the available data. This is done by minimizing the negative log-likelihood of
the data with respect to its marginal distribution (which is defined by the hyperparameters). 
For data $\bs{d}$, given the training configurations $T$ and hyperparameters $\bs{\alpha}$, the optimal hyperparameter 

\begin{equation}
    \bs{\alpha}^* = \underset{\bs{\alpha}}{\mathrm{argmin}} - \ln \mathbb{P}\LRs{\bs{d}|T,\bs{\alpha}}
    \label{eq:mle}
\end{equation}

with the Gaussian negative log-likelihood
given by

\begin{equation}
    -2 \ln \mathbb{P}\LRs{\bs{d}|T,\bs{\alpha}} = \snor{T} \ln 2\pi + \ln \det{K_{TT}} + \LRp{\bs{d}-\bs{m}}^{\T}\LRp{K_{TT}}^{-1}\LRp{\bs{d}-\bs{m}}
\end{equation}

where $\bs{m}$ is the prior mean prediction, $K_{TT}$ is the covariance matrix for the QoI evaluations over the training set $T$, and $|\cdot|$ denotes cardinality of the set. 

In practice, it is difficult to avoid local minima when optimizing \eqref{eq:mle}, so we employ multiple restarts with randomized initialization, taking the best results and rejecting those resulting in numerically singular or numerically zero results. Additionally, we observed the optimization of \eqref{eq:mle} sometimes resulted in posteriors with very small variance and extraneous oscillation. To counteract this, we introduced regularization terms to \eqref{eq:mle}, penalizing the resulting basis functions that have large negative parts or more oscillation than should be expected from the problem. This can be interpreted as a maximum a posteriori (MAP) estimation problem where we have assigned a prior to the hyperparameters, defined such that we expect them to produce bases with desirable characteristics.
Specifically, we postulate that the data-weighted basis functions
\begin{enumerate}
    \item attain their maxima close to 1.
    \item are neither Kronecker deltas, nor are close to 1 everywhere.
    \item have small negative mass.
\end{enumerate} 
For each, we design either gamma or normal densities and tune the mode to be a heuristically ``desirable'' value. If the $i$-th basis function corresponding to data $\LRp{\boldsymbol{\theta}_i,\boldsymbol{\xi}_i}$ is denoted as $f_i$, we define the specific forms as
\begin{align}
\ln \mathbb{P}_{\mathrm{peak}}\LRs{\boldsymbol{\alpha}} &=
  -\frac{1}{2\sigma_B^{2}}\sum_{i=1}^{\snor{T}}
   \LRp{\max\LRp{f_i} - 1}^{2} + c_{\mathrm{peak}},
  && \sigma_B = 0.2 \\
\ln \mathbb{P}_{\mathrm{norm}}\LRs{\boldsymbol{\alpha}} &=
  \sum_{i=1}^{\snor{T}}\LRp{\frac{\mu_n^{2}}{\sigma_n^{2}}\ln\|f_i\|
   - \frac{\mu_n}{\sigma_n^{2}}\|f_i\|} + c_{\mathrm{norm}},
  && \mu_n = \sqrt{\frac{\snor{\Theta \times \Xi}}{\snor{T}}},\;\; \sigma_n = 0.15 \\
\ln \mathbb{P}_{\mathrm{negative}}\LRs{\boldsymbol{\alpha}} &=
  -\frac{1}{2\sigma_\nu^{2}}\sum_{i=1}^{\snor{T}}\|\LRp{f_i}_{-}\|^{2} + c_{\mathrm{negative}},
  && \sigma_\nu = 0.01
\end{align}
with some constants $c$. For the first prior term, we model the maximum of the basis functions as independently normal with mean 1. For the second, the norms of the basis functions are assumed to be independently Gamma-distributed with a mode equal to the square-root of the parameter space measure divided by the number of training points. Note that we set the shape and scale parameters so that the mode is the desired value $\mu_n$ and the strength of the prior is given by a relative scale $\sigma_n$. The last prior assumes the norms of the negative parts of each basis function are independently (half) normal, with mode 0. The resulting log prior is given by the sum of these terms.
\begin{equation}
\begin{gathered}
    \ln \mathbb{P}\LRs{\bs{\alpha}}=\ln \mathbb{P}_{\mathrm{peak}}\LRs{\boldsymbol{\alpha}} + \ln \mathbb{P}_{\mathrm{norm}}\LRs{\boldsymbol{\alpha}} + \ln \mathbb{P}_{\mathrm{negative}}\LRs{\boldsymbol{\alpha}}
\end{gathered}
\end{equation}
Using Bayes' Rule, we write the log proportional posterior as
\begin{subequations}\label{eq:map}
    \begin{gather}
        \mathbb{P}\LRs{\bs{\alpha}|T,\bs{d}} \propto \mathbb{P}\LRs{\bs{d}|T,\bs{\alpha}} \mathbb{P}\LRs{\bs{\alpha}} \\
        \ln \mathbb{P}\LRs{\bs{\alpha}|T,\bs{d}} = \ln \mathbb{P}\LRs{\bs{d}|T,\bs{\alpha}} + \ln \mathbb{P}\LRs{\bs{\alpha}} + c
    \end{gather}
\end{subequations}
We then employ L-BFGS-B to minimize this (log) cost over the hyperparameters with multiple randomized restarts and rejection.

\section{Adaptive Sampling}\label{sec:adapt}
Since our problems of interest assume a prohibitively expensive HD solver, we are interested in regressing the
QoI response surface with as few HD samples as possible. It is difficult to determine how many samples are
sufficient a priori, much less the specific configurations to sample at. To this end, so-called \emph{adaptive
sampling} algorithms are of interest to expensive many query problems.

In these methods, the prediction is iteratively improved by only sampling at each step when needed. Given some
indicator of error or uncertainty, an acquisition function is used to identify the best place to acquire a new
sample if the current prediction is short of some tolerance. For standard GPR applications in the literature,
this acquisition function is typically taken to be the posterior variance, where the point with the greatest
posterior variance is sampled for the next iteration. The goal of this type of greedy approach is to find the next samples to reduce the variance.
When using stationary kernels, this typically corresponds to points in the domain far away from existing samples. In contrast, nonstationary kernels, such as our adjoint
inner product kernel, can cluster samples around points which ``need'' more information, as predicted by the
kernel. In our case, this would be where the LD solutions vary quickly, i.e. where they are sensitive to
changes in the design parameters.

Since the LD physics-aware kernel gives larger uncertainty where we should sample, we acquire the point of maximal variance at each iteration. For multiple quantities, we consider the maximizer of the pointwise variance sum, weighted by the precision of each measurement. This means we select the point which is the most uncertain, and favor quantities based on how precise their observations are. Specifically,

\begin{equation}
    (\bs{\theta},\bs{\xi})_{t+1} = \argmax_{\LRp{\bs{\theta}^*,\bs{\xi}^*} \in A_t \backslash T_t } \sum_n \frac{\mathrm{Var}\LRs{\,q^{(n)\dagger}\LRp{\bs{\theta}^*,\bs{\xi}^*}\,|\,d\LRp{T_t}}}{\sigma_{n}^{2}}
    \label{eq:aquisition_obj}
\end{equation}

where $A_t$ and $T_t$ are the acquirable set (we choose fixed, evenly spaced) and training set (i.e. initial samples and those previously chosen), respectively, at iteration $t$.
Here, $\sigma_{n}$ is the standard deviation of the measurement noise for the $n$-th QoI.
Another possibility is to replace the $n$-th noise variance by the $n$-th minimal pointwise variance, so that we favor observations with large spatial variation in their pointwise variance. There exist other acquisition functions from optimal experimental design, (e.g., A-optimality, I-optimality) whose incorporation we leave to future work. Our acquisition in \eqref{eq:aquisition_obj} is of the V-optimal (variance) type. 

\section{Neural Network Accelerated Kernel}\label{sec:nn}
\subsection{Framework}

We modify the GF algorithm to include an additional step before GP training where some set of the LD points is used to train a neural network (NN) for each QoI to provide

\[
\mathrm{Cov}\LRs{q^{\LRp{n}}\LRp{\bs{\theta}_i,\bs{\xi}_i},q^{\LRp{n}}\LRp{\bs{\theta}_j,\bs{\xi}_j}} \approx k_{NN}^{(n)}((\bs{\theta}_i,\bs{\xi}_i), (\bs{\theta}_j,\bs{\xi}_j))
\]

without requiring any additional forward and adjoint LD solves during online inference, as would be required for every new parameter when evaluating \eqref{eq:four_kernels} directly.
In the present work, we only consider learning a covariance where QoIs are a priori independent.
This is advantageous when fast online inference is desirable but the desired set of test parameters should be large or only its distribution is known at training time, i.e. in order to provide a reduction in total cost and to provide reliable evaluations.

Since we wish to approximate a known kernel, we have to carefully ensure the approach can learn arbitrary symmetric positive definite kernels.
This is trickier than parameterizing a symmetric positive definite matrix of known size, such as done in existing literature \cite{wilson2016deep}, as we wish to evaluate at any arbitrary parameters, even those not known at training time. 

We propose a novel NN approach for approximating an arbitrary kernel to speed up evaluation through approximating a finite sequence of functions $\bs{\phi}_1,\dots,\bs{\phi}_M: \mbb{R}^P \rightarrow \mbb{R}$.
Additionally, learnable log-weights $\tau_1, \dots, \tau_M$ are used to ease the training of smaller contributions.
Specifically, we consider the neural network of the following form

\begin{equation}
    k_{NN}^{(n)}((\bs{\theta}_i,\bs{\xi}_i), (\bs{\theta}_j,\bs{\xi}_j)) :=
    \sum_{m=1}^M e^{2\tau_m} \bs{\phi}_m^{(n)}((\bs{\theta}_i,\bs{\xi}_i)) \bs{\phi}_m^{(n)}((\bs{\theta}_j,\bs{\xi}_j))
\label{eq:nn_kernel}
\end{equation}

This specific form is motivated by Mercer's theorem~\cite{mercer1909xvi}.
Mercer guarantees the existence of a sequence $\LRc{\sqrt{\lambda_i}\psi_i(\bs{x})}$ where $\psi_i$ are orthonormal and continuous and give the series expansion of a positive definite kernel (Mercer kernel)

\begin{equation}
    k(\bs{x},\bs{x}') = \sum_{i=1}^{\infty} \lambda_i \psi_i(\bs{x}) \psi_i(\bs{x}')
\end{equation}

with uniform and absolute convergence.
That is, the neural network universal approximation theorem (see, e.g., \cite{bui2023} and references therein) for each $\psi$ combined with Mercer's theorem gives universal approximation for NNs of form \eqref{eq:nn_kernel} and any Mercer kernel.
However, $\psi_i$ from the canonical Mercer series are eigenfunctions of the Hilbert-Schmidt operator associated with the kernel. We do not directly seek these or enforce orthonormality. Rather, we train some $\bs{\phi}_1,\bs{\phi}_2,...,\bs{\phi}_M$ that minimize some loss involving the finite term approximation.

\subsection{Training}

We use a minibatched elementwise least squares loss, using the upper triangular part to avoid overcounting symmetric components. Specifically, the loss function

\begin{subequations}
    \begin{gather}
        \mathcal{J}_{\text{mini}} := \frac{1}{|I_\text{mini}|} \sum_{(i,j) \in I_\text{mini} \subset I_\text{train}} \LRp{\LRs{K}_{i,j} - \LRs{K_{NN}}_{i,j}}^2 
        \label{eq:nn_loss} \\
        \LRs{K}_{i,j}:=k\LRp{(\bs{\theta}_i,\bs{\xi}_i), (\bs{\theta}_j,\bs{\xi}_j)} \\
        \LRs{K_{NN}}_{i,j}:=k_{NN}\LRp{(\bs{\theta}_i,\bs{\xi}_i), (\bs{\theta}_j,\bs{\xi}_j)}
    \end{gather}
\end{subequations}

where $I$ are multi-index sets. The subscript ``mini'' refers to the current minibatch and the square brackets and subscripts are used to denote taking the $\LRp{i,j}$-th element of the matrix in brackets.
After selecting $N_{\text{train}}$ parameters for training, we have the training index set

\begin{align}
    I_\text{train} :=
    \biggl\{ (i,j)\in\{1, \dots, N_{\text{train}}\}^2 : j \geq i \biggr\}
\end{align}

The minibatch approach involves performing gradient descent steps based on the gradient of $\mathcal{J}_{\text{mini}}$, rather than the loss over the entire training dataset. We partition $I_{\text{train}}$ into subsets $I_{\text{mini}}$ and iterate over them until some number of ``epochs'' (times covering all of $I_{\text{train}}$) has passed.

Before splitting the data into test and train subsets, it is desirable to shift and scale the dataset to near a unit domain and range, as is standard practice in machine learning. 
A common choice is to standardize the inputs and outputs (e.g., by enforcing zero mean and unit standard deviation, or by mapping into a unit hypercube), but our dataset consists of PSD matrices generated by Mercer kernels, which are not closed under element-wise sum with arbitrary matrices.
However, we can find the largest scalar, within numerical precision, that can be subtracted and retain positive definiteness. In our case, we thus consider a unit range to refer to a restriction on the cone of positive semi-definite matrices such that the maximal element has unit magnitude.
To that end, let $K$ denote the $N \times N$ PSD kernel matrix constructed from the train and test data.
And let $\lambda_1(M)\geq\dots\geq\lambda_N(M)$ be the eigenvalues of a given $N\times N$ matrix $M$.
Consider the shifted matrix

\begin{equation}
    \tilde{K} := K - \mu \mb{1} \mb{1}^\T,\quad \mu>0
\end{equation}

where $\bs{1}$ is a vector of all ones. Then Weyl's inequality applied to $\tilde{K} + \mu \mb{1} \mb{1}^\T$ gives
\[
\lambda_i(\tilde{K}) + \lambda_j(\mu \mb{1} \mb{1}^\T) \leq \lambda_{i+j-N}(K)
\]
Apply again to $K-\mu \mb{1} \mb{1}^\T$ and we have $$\lambda_i(K) + \lambda_j(-\mu \mb{1} \mb{1}^\T) \leq \lambda_{i+j-N}(\tilde{K}).$$
Exploiting the $N-1$ eigenvalues equal to $0$ in $\mu \mb{1} \mb{1}^\T$ by setting $j=N$ in the first relation and $j=N-1$ in the second while considering $i$ one higher gives
\[
\lambda_{i+1}(K) \leq \lambda_i(\tilde{K}) \leq \lambda_i(K)
\]
Through considering $i=N-1$, we now know that the second smallest eigenvalue of $\tilde{K}$ is larger than the smallest eigenvalue of $K$, and repeating this for different $i$ yields $$\lambda_1(\tilde{K})...\lambda_{N-1}(\tilde{K}) \geq \lambda_N({K}) \geq 0.$$ 
Consequently, we seek the critical value $\mu_\text{crit}$ where $\lambda_N(\tilde{K})=0$, and then we choose $\mu$ as

\begin{equation}
    \mu := \LRp{1-\varepsilon_m}\mu_{\text{crit}}
\end{equation}

where $\varepsilon_m$ is the floating point precision (i.e., $2^{-52}$ in the IEEE 754 standard 64-bit architecture).
Since the determinant is the product of eigenvalues, and all other eigenvalues are greater than 0, we can find $\mu_\text{crit}$ by solving for $\det(\tilde{K})=0$.
Applying the matrix determinant lemma \cite{ding2007det} gives

\[
    \det(\tilde{K})=(1 - \mu \mb{1} K^{-1} \mb{1}^\T )\det(K)
\]    

and finally we have 

\begin{equation}
    \mu_\text{crit} = 1/(\mb{1} K^{-1} \mb{1}^\T)
\end{equation}

In practice, we compute the shift via a Cholesky factorization of $K$ followed by a sum reduction. Finally, after shifting, the data is rescaled by dividing by the maximal element of $\tilde{K}$.

\subsection{Evaluation}

After training, we use the NN to construct $N_{\text{train}} \times N_{\text{test}}$ matrices of kernel evaluations evaluated at the Cartesian product of parameter vectors for use in the computations of the GPR posterior (where  $N_{\text{test}} \geq N_{\text{train}}$). It is efficient and convenient to evaluate the network at $N_{\text{train}}+N_{\text{test}}$ inputs, or just $N_{\text{test}}$ if parameter vectors are equal such as for the marginal covariances for the train and test sets, producing all the needed $\bs{\phi}_1, \dots, \bs{\phi}_M$ evaluations. In popular linear algebra libraries \cite{harris2020array, jax2018github}, the final kernel computation can be accomplished concisely using broadcasting of the series basis and coefficients followed by a sum reduction. In practice, the dominant cost will be evaluating the neural network, i.e. the evaluation scales with $\mathcal{O}(N_{\text{test}})$. 

\section{Algorithm}\label{sec:algorithm}

The overall inference and data acquisition process is summarized in Algorithm \ref{algo:overall}, using the evaluation of our kernel as described in Algorithm \ref{algo:gpr}.
The additional step of pre-correcting the LD QoIs using the empirical statistics of the HD observations to shift and scale as described in Algorithm \ref{algo:overall} is motivated by the superior performance and can be reconciled with the theory by considering a modified LD observation functional, where the scaling does effect the adjoint solution up to a factor but ultimately can be canceled from the trained scaling factors in the kernel parameterization.
That is, it is justified in practice to just scale and shift the LD QoI without modifying the numerics of the observation functional and adjoint.

\begin{algorithm}
\caption{Gaussian Functional Regression}
\begin{algorithmic}[1]
\REQUIRE HD parameters $\{ (\theta,\xi) \in T_t \}$ and corresponding $q_H$ at each for GP training, and LD parameters $\{ (\theta,\xi) \in  A_t \}$ for kernel samples
    \STATE \textbf{Gaussian Functional Regression}
    \IF {NN is trained}
        \STATE \hspace{6pt} Evaluate kernel at all training and desired sample locations by plugging parameters into Eqn~\eqref{eq:nn_kernel}
    \ELSE
        \STATE \hspace{6pt} Run LD simulations at all parameters, obtaining solutions and any needed adjoints
        \STATE \hspace{6pt} Evaluate kernel at all training and desired sample locations by plugging parameters as well as LD solutions and/or adjoints into Eqn~\eqref{eq:four_kernels}
    \ENDIF
    \STATE Obtain posterior mean, Eqn~\eqref{eq:gp_post_mean}
    \STATE Obtain posterior covariance, Eqn~\eqref{eq:gp_post_cov}
\end{algorithmic}
\label{algo:gpr}
\end{algorithm}

\begin{algorithm}
\caption{GFR with training acquisition for a single QoI}
\begin{algorithmic}[1]
    \STATE \textbf{Hyperparameter Optimization}
    \STATE Use empirical mean and standard deviation of available HD QoI samples to shift and rescale $q_L$ as the prior mean
    \STATE Maximize Log-Marginal Likelihood \eqref{eq:mle} or Log-Posterior \eqref{eq:map} via gradient based methods to provide kernel hyperparameters

    \STATE \vspace{0.2cm} \textbf{Neural Network Training (Optional)}
    \STATE Shift dataset (kernel samples) by $\LRp{1-\varepsilon_m}\mu_\text{crit}$ and scale by maximal element
    \STATE Remove some elements from $I_\text{train}$ to save as test set
    \STATE Train NN until converged using Eqn~\eqref{eq:nn_loss}
    
    \STATE \vspace{0.2cm} \textbf{Gaussian Functional Prediction}
    \STATE Use current HD samples to get posterior QoI response via Algorithm~\ref{algo:gpr}
    
    \STATE \vspace{0.2cm} \textbf{Acquisition \& Sampling (Optional)}
    \STATE Find the next query point, maximizing Eqn~\eqref{eq:aquisition_obj} based on posterior variance
    \STATE Run the HD model to acquire new data
    \STATE Continue with new Hyperparameter Optimization unless maximum iterations reached
\end{algorithmic}
\label{algo:overall}
\end{algorithm}

\section{Results}\label{sec:results}
\subsection{Test Problem: Winglet Geometry}

For our test problem of interest, we correct predictions of the aerodynamic coefficients provided by a 2D simulation over an airfoil to better model the ground-truth values given by a 3D simulation over the full wing. For the governing equations, we use the compressible Reynolds-Averaged Navier-Stokes (RANS) equations. 
For the turbulence model, we use the 2003 Menter Shear Stress Transport model (SST-2003) \cite{menter2003sst}. This is a two-equation linear eddy
viscosity model, which introduces conservation equations for the turbulent kinetic energy $k$ and specific dissipation rate $\omega$. We refer readers 
to Menter and related works for the full closure definition. This results in solutions with six-vector values in 3D and five-vector values in 2D. In 
conservative form, these components are density, momentum, energy, and the turbulence quantities. We use SU2 \cite{economon2016su2} as the primary and
adjoint solver. We review the laminar compressible Navier-Stokes equations

\begin{subequations}    \label{eq:cns}
    \begin{gather}
        \frac{\partial U}{\partial t} + \nabla \cdot F_{c}\!\LRp{U} - \nabla \cdot F_{v}\!\LRp{U,\nabla U} - S = 0 \\
        U = \LRs{\rho, \rho u, \rho v, \rho w, \rho E}^{\T} \\
        F_{c} = \begin{bmatrix}
            \rho \bs{u} \\
            \rho \bs{u} \otimes \bs{u} + pI \\
            \rho E \bs{u} + p \bs{u}
        \end{bmatrix} \\
        F_{v} = \begin{bmatrix}
            \bs{0} \\
            \tau \\
            \tau \cdot \bs{u} + \kappa \nabla T
        \end{bmatrix} \\
        \tau = \mu \LRp{\nabla \bs{u} + \nabla \bs{u}^{\T}} - \frac{2}{3}\mu I\LRp{\nabla \cdot \bs{u}}
    \end{gather}
\end{subequations}

where $\rho$ is the density, $u$, $v$, and $w$ are components of the velocity $\bs{u}$, $E$ is the specific energy, $p$ is the pressure, and $T$ is the
temperature. The modeled terms include the viscous stress tensor $\tau$, the viscosity $\mu$, and the thermal conductivity $\kappa$. For concision,
we use $F_c$ and $F_v$ as the convective and viscous flux tensors, respectively, $S$ is a forcing term, and $I$ is the $3 \times 3$ identity matrix. 

The RANS equations are derived by time-averaging \eqref{eq:cns} to obtain a mean flow solution $\bar{U}$ and modeling the terms that arise as functions of
the deviatoric quantities $\hat{U}$. Linear eddy viscosity models, including the SST-2003 model we use, assume the so-called Reynolds stress $\tau_t$
is linearly related to the deviatoric strain rates through an eddy viscosity term $\mu_t$ (the Boussinesq assumption):

\begin{equation}
    \tau_t = \mu_t \LRp{\nabla \hat{\bs{u}} + \nabla \hat{\bs{u}}^{\T}} - \frac{2}{3} \mu_t I \LRp{\nabla \cdot \hat{\bs{u}}}
\end{equation}

The 3D geometry is based on the ONERA M6 wing \cite{schmitt1979pressure,destarac2016onera}, which is a common validation case for turbulent external
flows. We have designed a continuous blended winglet geometry parameterize by $\xi$, which is the height of the wingtip chord above the ONERA M6 chord plane
$h$ normalized by the root chord length $C$, i.e.,

\begin{equation}
    \xi = h / C
\end{equation}

\begin{figure*}
    \centering
    \begin{tabular}{cc}
        \STAB{\includegraphics[width=0.5\textwidth]{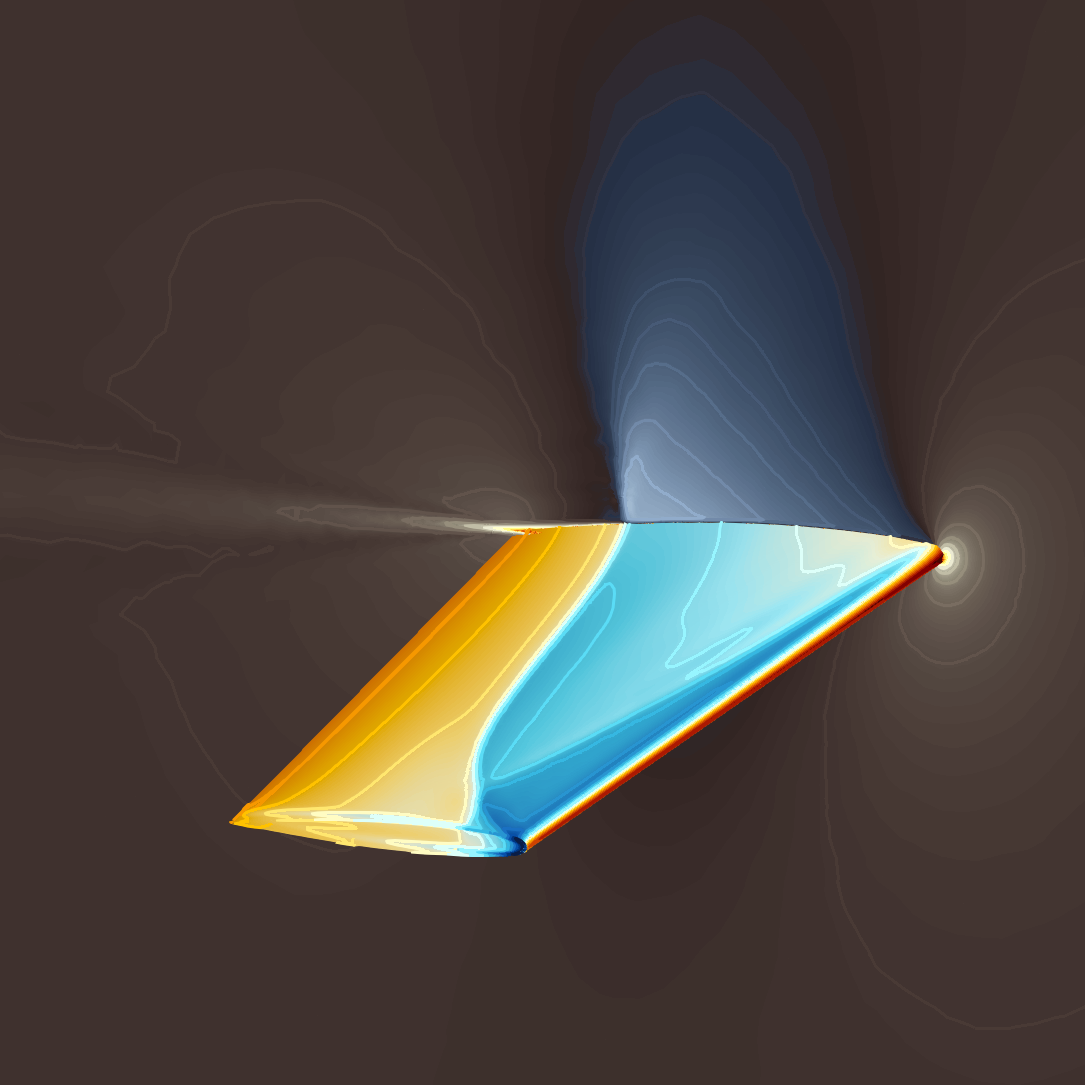}} & \STAB{\includegraphics[width=0.16\textwidth]{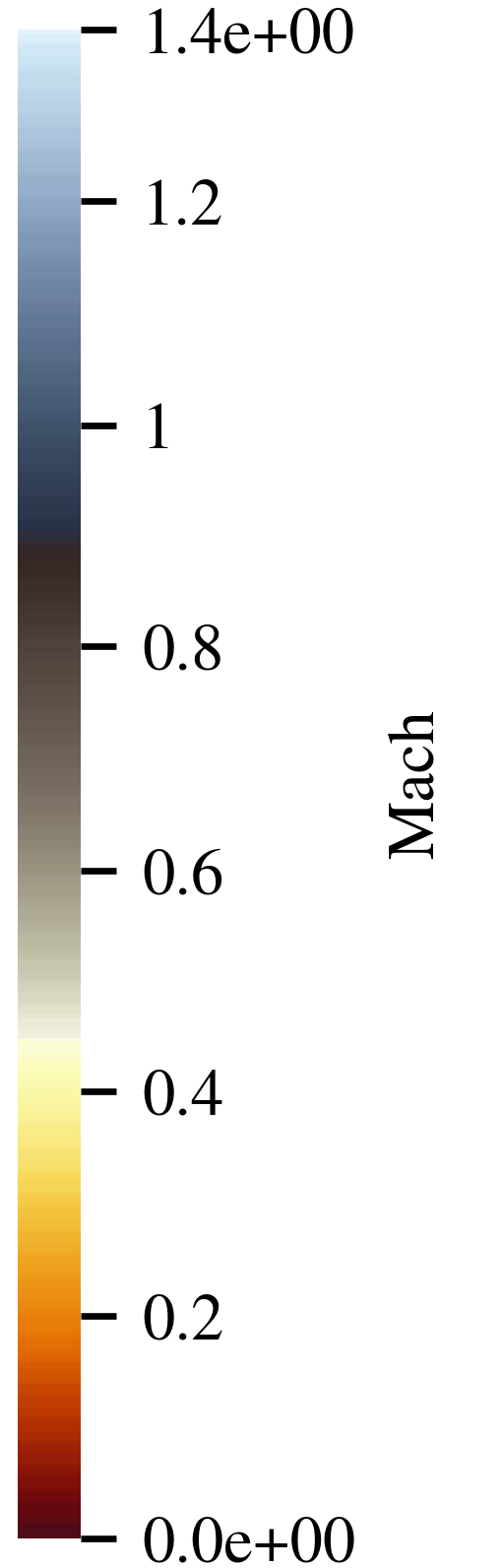}}\\
        \STAB{\includegraphics[width=0.5\textwidth]{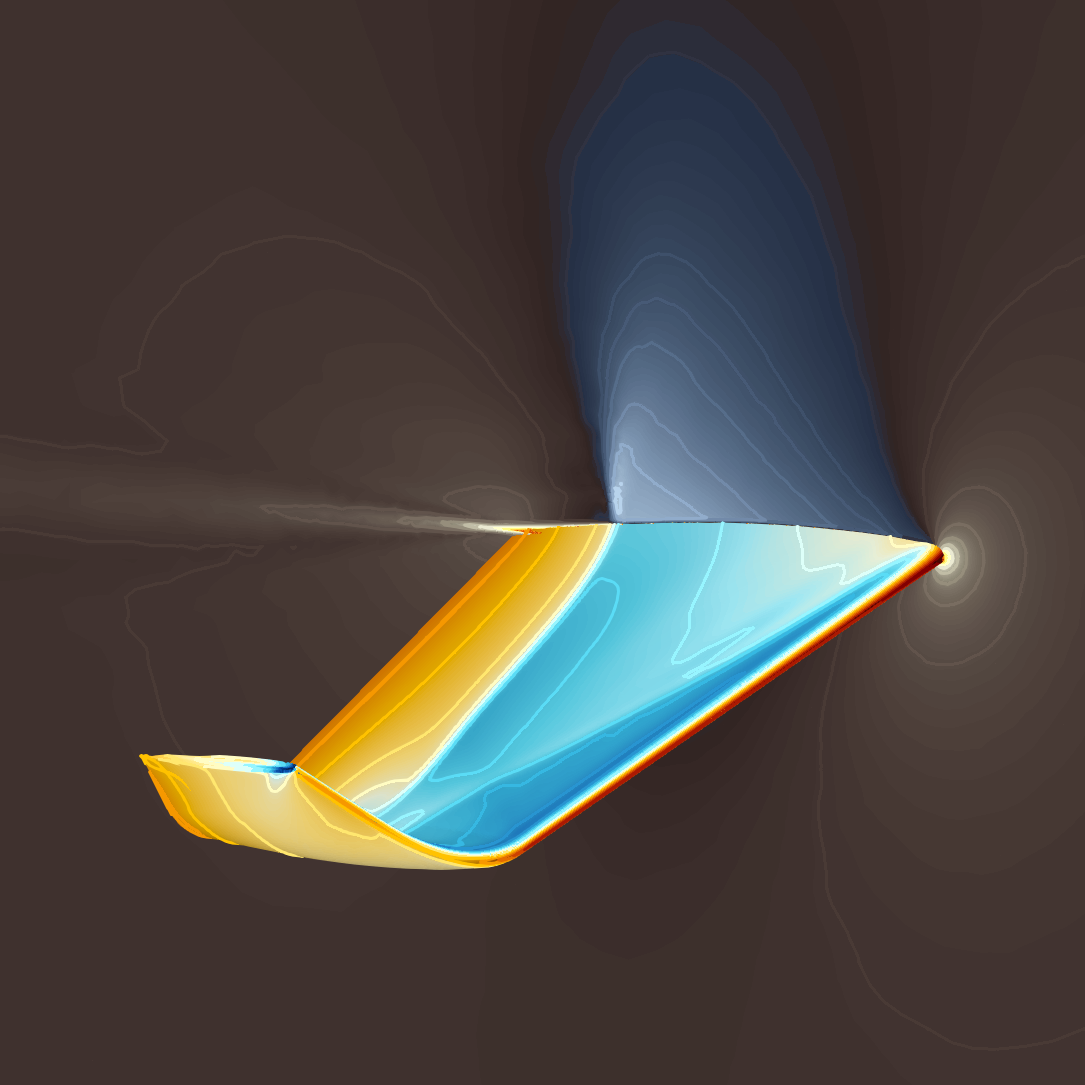}} &
        \STAB{\includegraphics[width=0.16\textwidth]{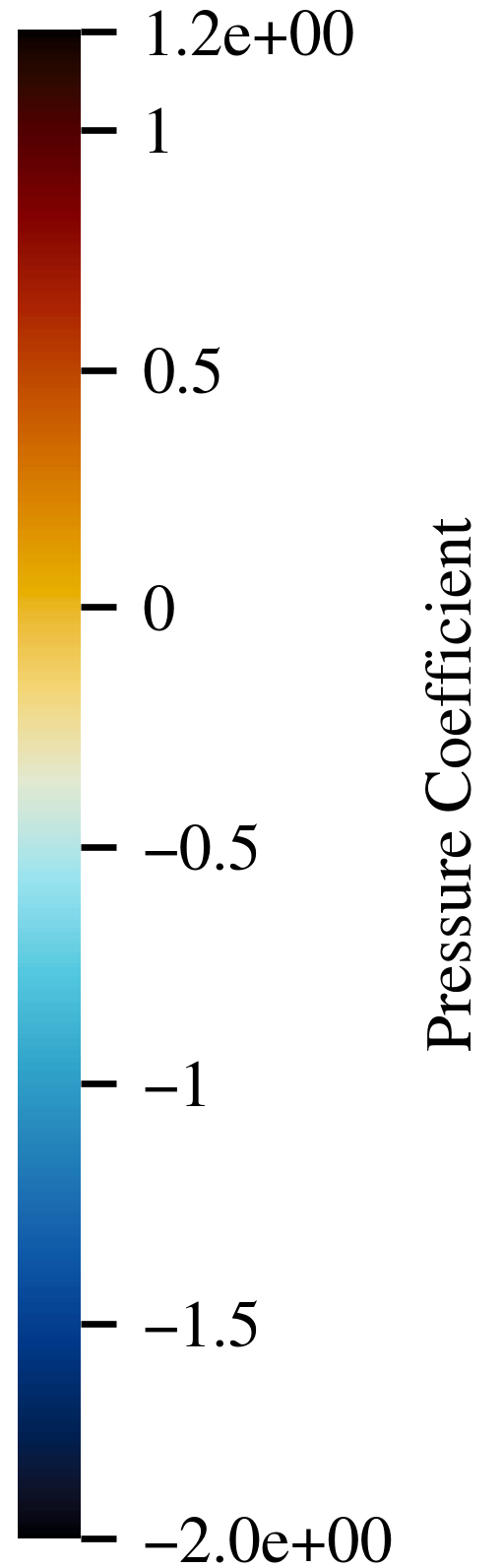}}
    \end{tabular}
    \caption{SU2 solutions for (Top) $\xi=0$ and (Bottom) $\xi=0.25$. The root plane and wing surface are colored by the Mach and pressure, respectively}
    \label{fig:ex_sols}
\end{figure*}

If  $\xi$ corresponds to a height at which the winglet has not reached the full $100^{\circ}$ cant, the geometry is truncated by a plane at the
current cant angle. If $\xi$ corresponds to a height after the winglet has reached the full cant angle, the winglet taper rate is decreased and the sweep operation is extended along the canted chord plane so that 
the wingtip size reaches a minimum at $0.281C$. We choose this wingtip size to match the same taper rate as the ONERA M6 until full cant is
reached. This results in a continuous variation in geometry, but creates two regions of changing behavior: (1) after the winglet reaches full cant,
the entire geometry changes past the original wingtip, whereas only the winglet tip grows before reaching full cant, and (2) when $\xi$ is zero, the
definition of $\xi$ being associated with a quantity normal to direction of geometry addition results in an infinite rate of change. Example solutions at a Mach number of 0.84, Reynolds number of 11.72 thousand, angle of
attack of 3 degrees, and $\xi$ of 0 and 0.25 are shown in Figure \ref{fig:ex_sols}.

Since the root airfoil does not change with $\xi$, we introduce an auxiliary solve around the winglet tip airfoil for when regressing across changes in $\xi$. The domain is taken to be the plane
that contains the wingtip. This plane starts as a span-normal plane at $\xi=0$ and cants to $100^\circ$ as $\xi$ increases. The freestream condition
is then projected onto the in-plane components and used as the 2D freestream condition, i.e.,

\begin{subequations}
    \begin{gather}
        \bs{u}_{2D}^{\LRp{\infty}} = \LRp{I-\bs{n}_{2D}\bs{n}_{2D}^{\T}}\bs{u}_{3D}^{\LRp{\infty}} \\
        \alpha_{2D} = \arctan\LRp{\frac{\nor{\LRp{I-\boldsymbol{e}_1\boldsymbol{e}_1^{\T}}\boldsymbol{u}_{2D}^{(\infty)}}}{\boldsymbol{e}_1\boldsymbol{e}_1^{\T}\boldsymbol{u}_{2D}^{(\infty)}}} \\
        \mathrm{Re}_{2D} = \frac{\nor{\boldsymbol{u}_{2D}^{(\infty)}}C_{tip}}{\nor{\boldsymbol{u}_{3D}^{(\infty)}}C_{root}}\mathrm{Re}_{3D} \\
        \mathrm{Ma}_{2D} = \frac{\nor{\boldsymbol{u}_{2D}^{(\infty)}}}{\nor{\boldsymbol{u}_{3D}^{(\infty)}}}\mathrm{Ma}_{3D}
    \end{gather}
\end{subequations}

where the infinity superscript denotes a freestream quantity, $\bs{n}_{2D}$ is the wingtip normal vector, $\bs{e}$ is the canonical basis vector,
$\alpha$ is the angle of attack, and $\mathrm{Ma}$ and $\mathrm{Re}$ are the Mach and Reynolds numbers, respectively.

\begin{figure*}
    \centering
    \includegraphics[width=\textwidth]{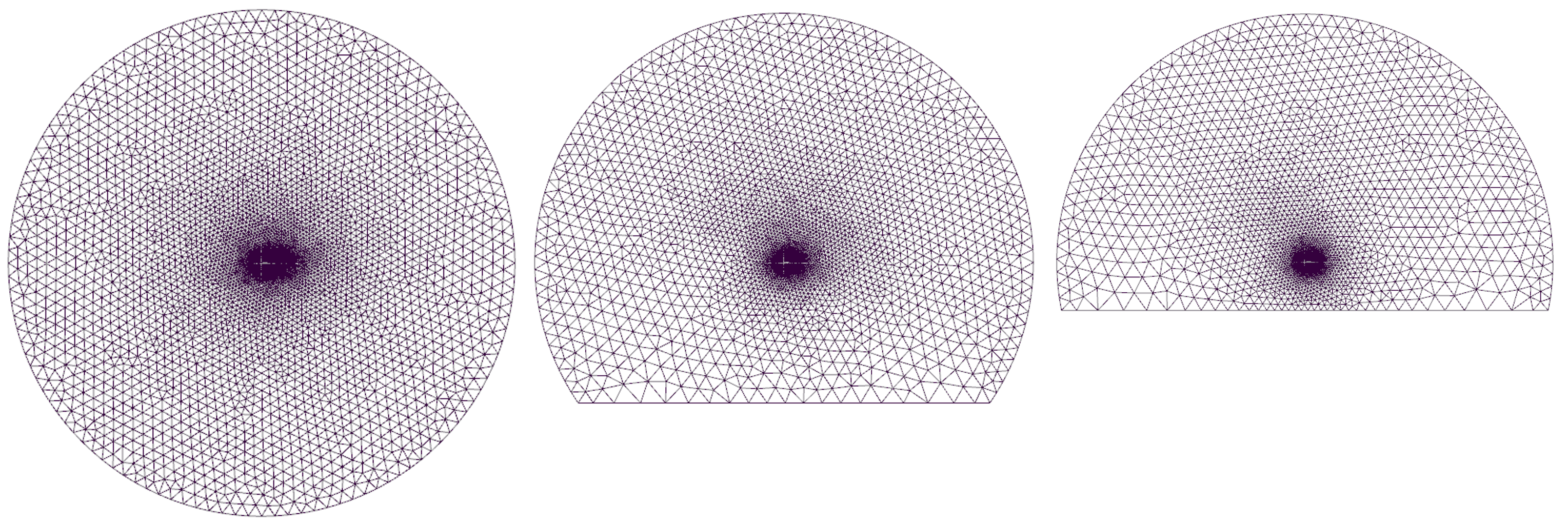}
    \caption{Auxiliary 2D meshes for (Left) $\xi=0$, (Center) $\xi=0.025$ and (Right) $\xi=0.125$.}
    \label{fig:ld_aux_mesh}
\end{figure*}

Since the 3D domain uses the symmetric boundary condition on the root plane, we use a slip boundary condition on the 2D wall. Examples of 2D meshes are
given in Figure \ref{fig:ld_aux_mesh}.

For the stationary kernels in the following examples, we use a mixture of Mat\'ern kernels, with regularity parameters $\nu=1/2,3/2,5/2,\infty$.
This allows the regularity of the regression to be inferred from data.

\subsubsection{Varying Winglet Size}

We evaluate the benefits of PDE-informed nonstationarity in the auxiliary kernel by comparing the product kernel involving the auxiliary primal solution (but not the primary LD solutions) to stationary
kernels while varying the winglet size. Since the winglet size does not impact the root plane, this regression relies solely on the auxiliary
kernels, i.e. it only uses the 2D wing tip solutions. As discussed previously, the design of the parameterization results in an infinite rate of change when $\xi=0$, which will give the
auxiliary solution kernel an opportunity to guide regression by increasing basis variability and posterior uncertainty close to $\xi=0$. The angle of attack is fixed to 3 degrees, the Mach number is fixed to 0.84, and the Reynolds number is fixed to
11.72 thousand.

\begin{figure*}[htp]
    \centering
    \includegraphics[width=\textwidth]{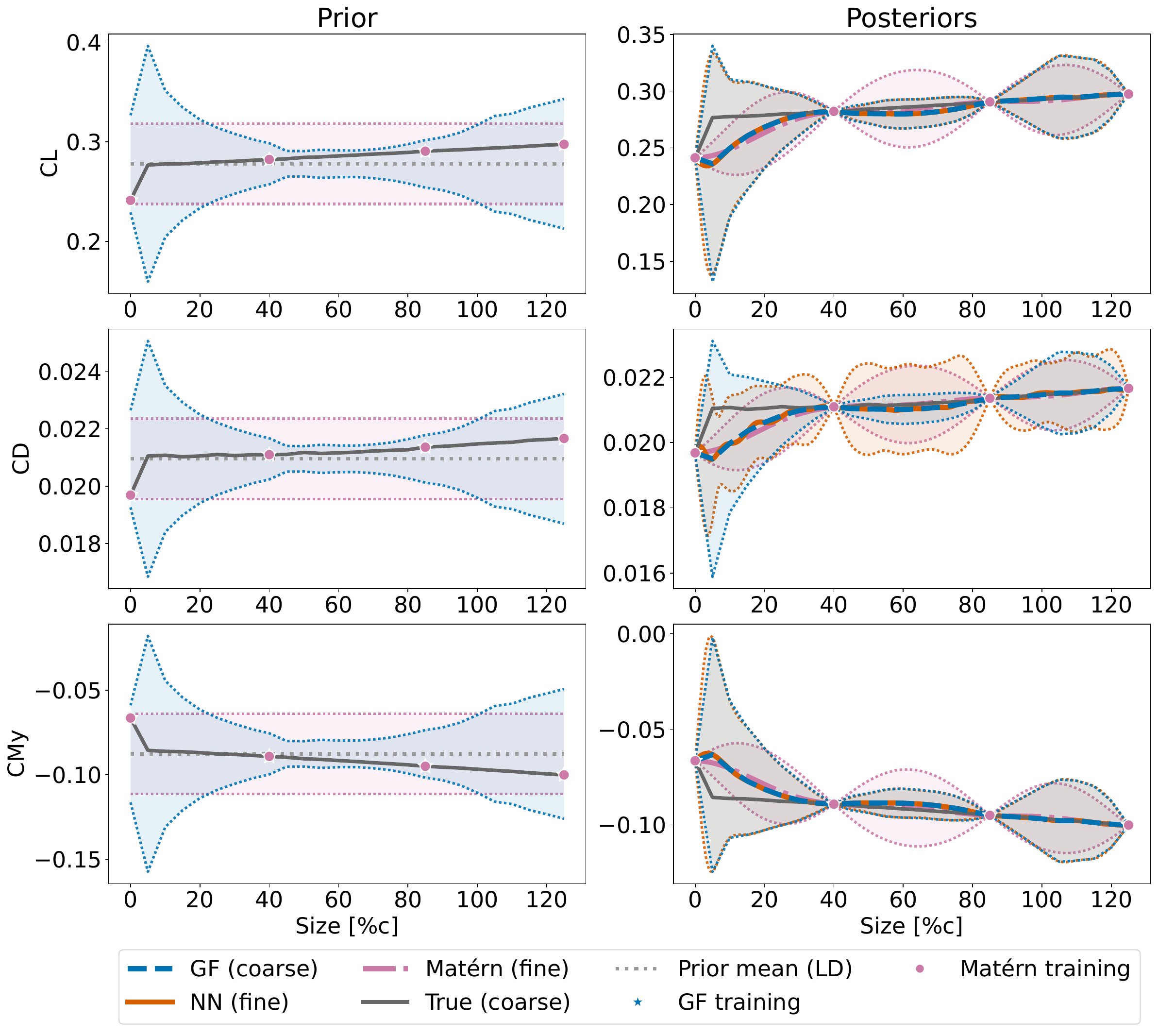}
        
    \caption{Comparison between our GF kernel and Mat\'ern mixture on the QoI regression at iteration 0 (before sampling).}
    \label{fig:adapt_0}
\end{figure*}

\begin{figure*}[htp]
    \centering
    \includegraphics[width=\textwidth]{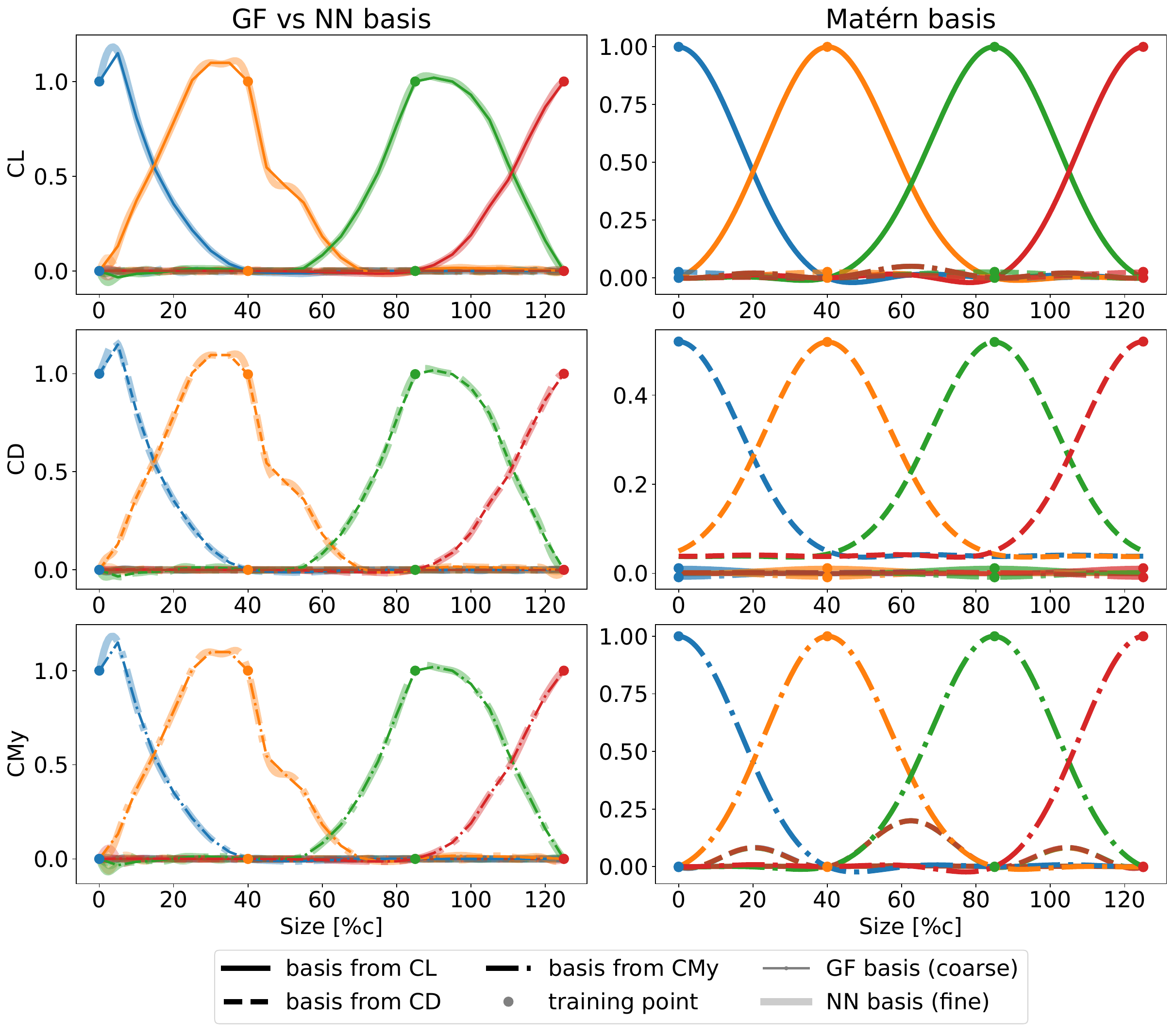}    
    \caption{Comparison between our GF kernel and Mat\'ern mixture basis functions at iteration 0 (before sampling).}
    \label{fig:adapt_0_bases}
\end{figure*}

\begin{figure*}[htp]
    \centering
    \includegraphics[width=\textwidth]{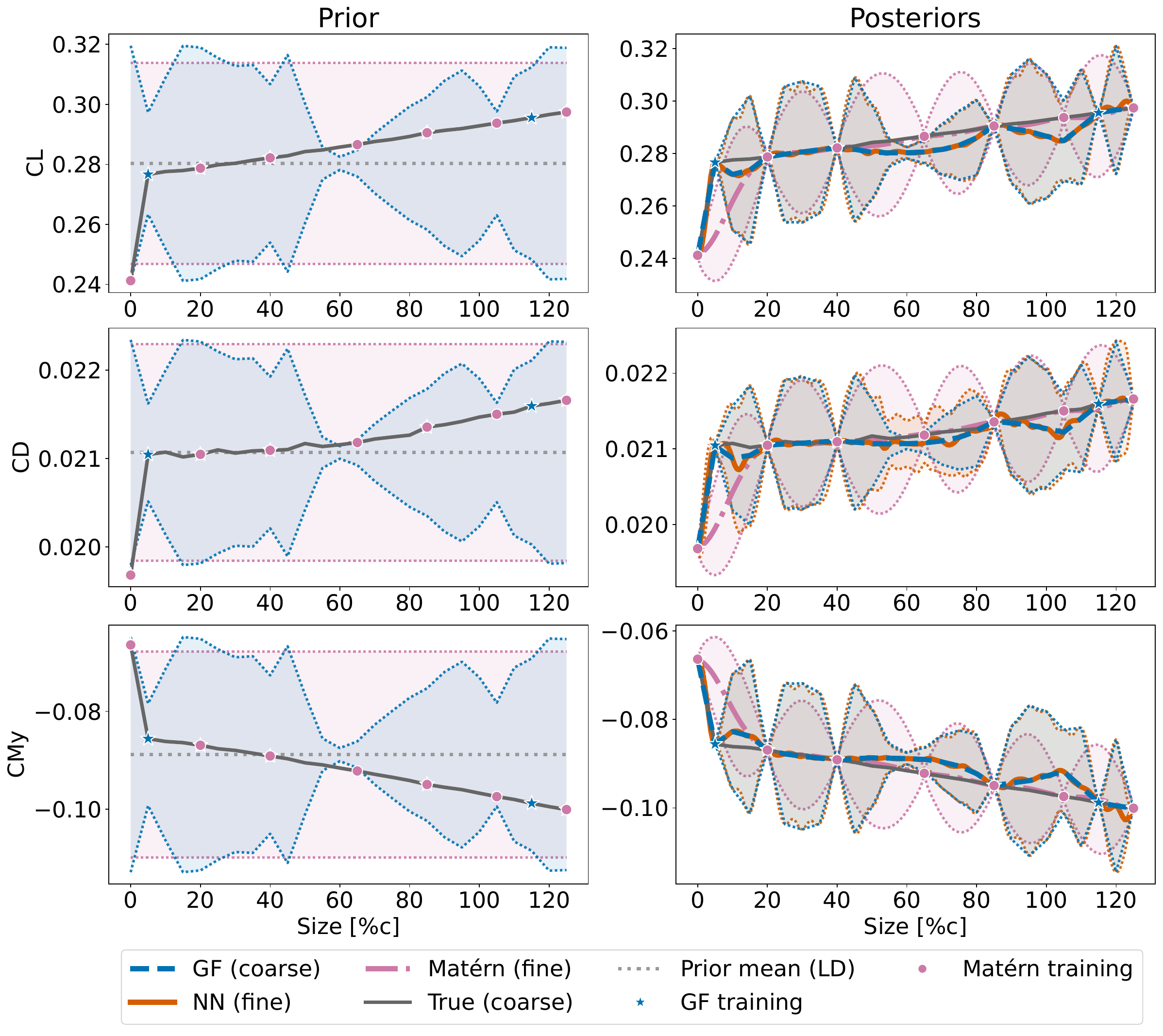}
        
    \caption{Comparison between our GF kernel and Mat\'ern mixture on the QoI regression at iteration 3.}
    \label{fig:adapt_3}
\end{figure*}

\begin{figure*}[htp]
    \centering
    \includegraphics[width=\textwidth]{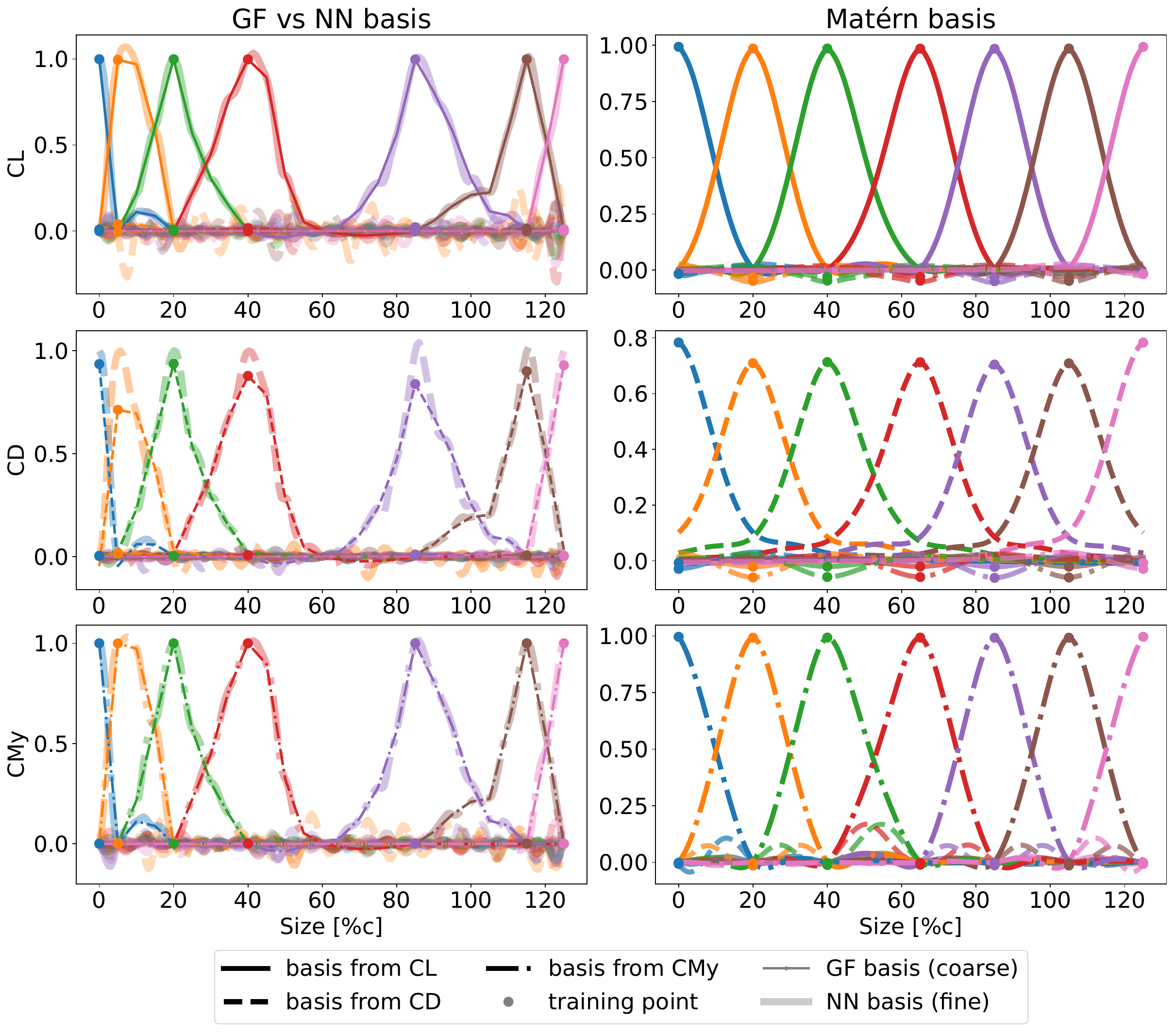}    
    \caption{Comparison between our GF kernel and Mat\'ern mixture basis functions at iteration 3.}
    \label{fig:adapt_3_bases}
\end{figure*}

To demonstrate the effectiveness of using the nonstationary kernel in an adaptive setting, we run an adaptive sampling algorithm using the two kernels to acquire the
point with the most posterior variance at each iteration, starting with 4 points in the domain, in Figures~\ref{fig:adapt_0} and \ref{fig:adapt_0_bases}. Regressions and bases at iteration 3 are  presented in Figures~\ref{fig:adapt_3} and \ref{fig:adapt_3_bases}, respectively.
As we acquire more samples, the PDE kernels 
drive acquisition near the regions of high uncertainty (large adjoint sensitivity), quickly resolving the features of interest. By comparison, the stationary kernels are unaware of the surface features until enough samples have been acquired to make shorter length scales more likely.

\subsubsection{Varying Flight Condition}

Additionally, we evaluate the benefits of the adjoint inner product kernel by comparing the regression with no winglet ($\xi=0$) across the so-called
critical Mach number, which is the Mach number at which the upper surface shock appears, resulting in rapid changes in aerodynamic forcing as
a function of Mach number.
By varying the flight condition, all changes are seen by the 2D root LD model and thus we use its primal and adjoint solutions in the kernel but not the tip solutions that the preceding varying winglet size results relied on.
Since the 2D model should also exhibit critical Mach behavior, we expect the
regression with the adjoint inner product kernel to allow for rapid change around this number. However, since such behavior is associated with
small length scales, a stationary kernel will not be able to resolve this behavior without observing many samples.

The regressions are given a grid covering an interval which includes the critical Mach number. The angle of attack is fixed at 3 degrees, and the Reynolds number is fixed to 11.72 thousand. We see that
the stationary kernel attempts to predict a curve with uniform variability, whereas the adjoint inner product kernel shows both (1) a rapid
change around the critical Mach number and (2) more slowly varying behavior before the critical Mach number.

Since the adjoint
inner product kernel is warping a stationary kernel, the uncertainty is able to drive sampling close to the critical Mach number, which is completely unseen by the stationary kernel alone.

\begin{figure*}[htp]
    \centering
    \includegraphics[width=\textwidth]{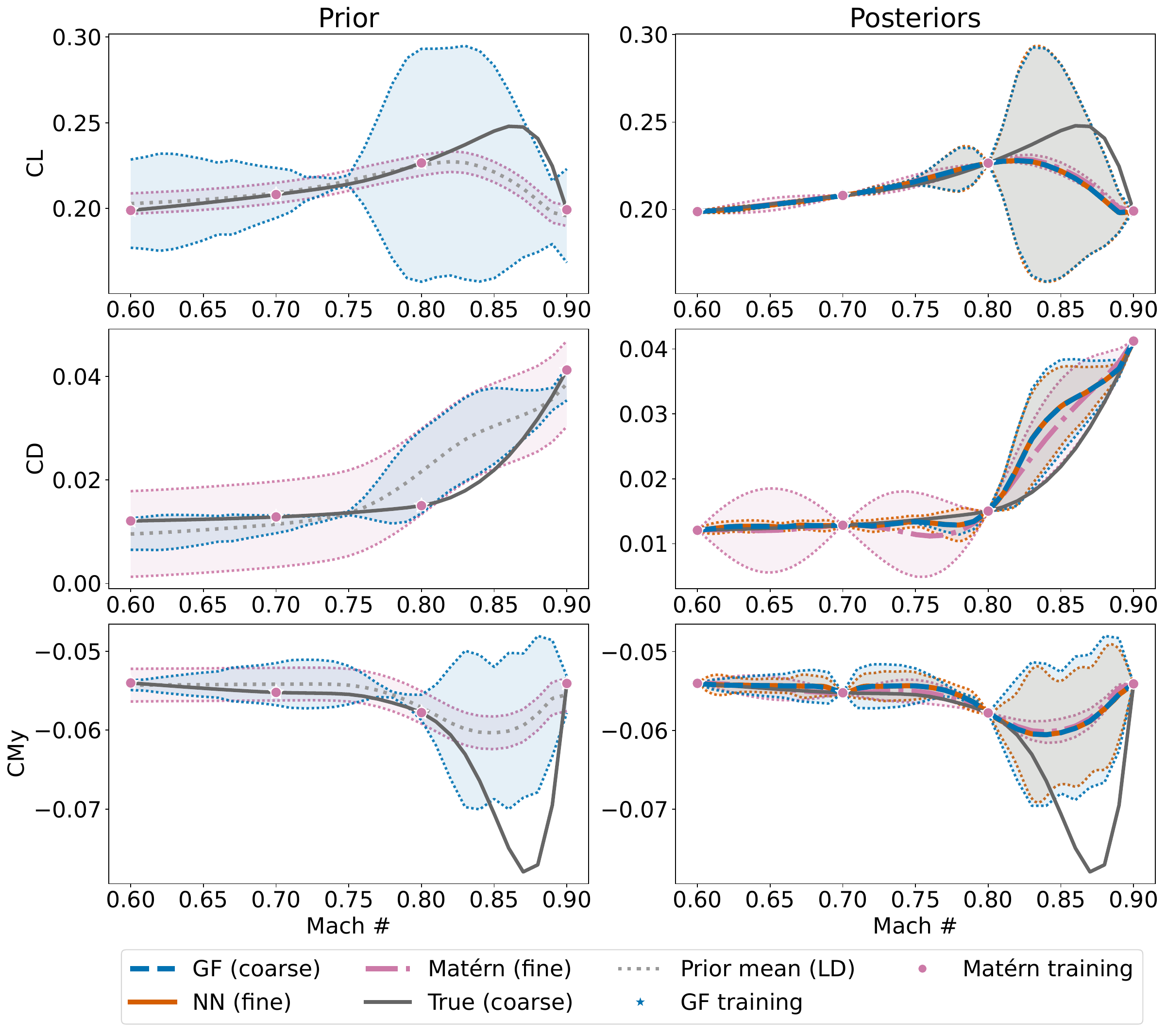}
        
    \caption{Comparison between our GF kernel and Mat\'ern mixture on the QoI regression at iteration 0 (before sampling).}
    \label{fig:mach_adapt_0}
\end{figure*}

\begin{figure*}[htp]
    \centering
    \includegraphics[width=\textwidth]{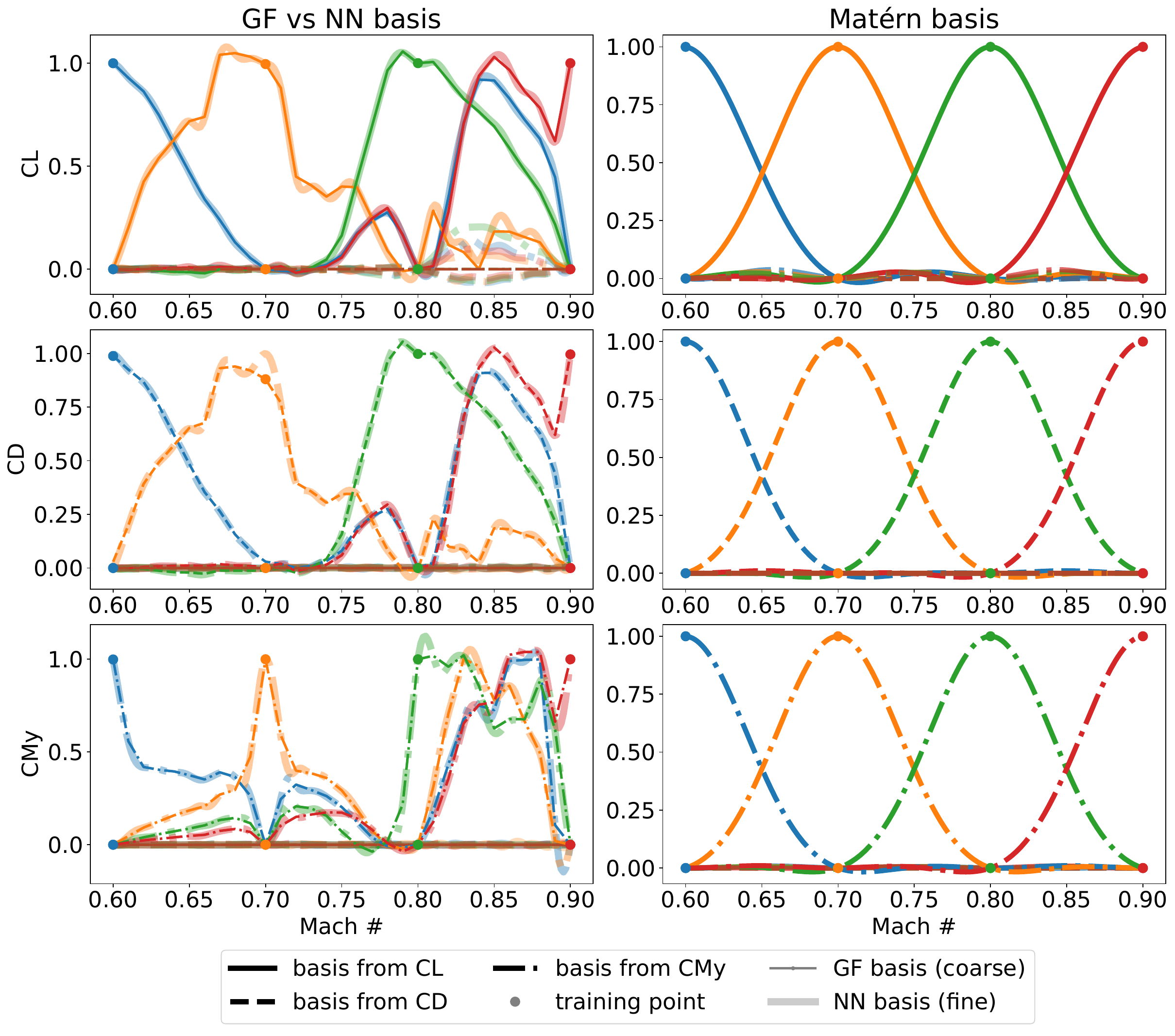}    
    \caption{Comparison between our GF kernel and Mat\'ern mixture basis functions at iteration 0 (before sampling).}
    \label{fig:mach_adapt_0_bases}
\end{figure*}

\begin{figure*}[htp]
    \centering
    \includegraphics[width=\textwidth]{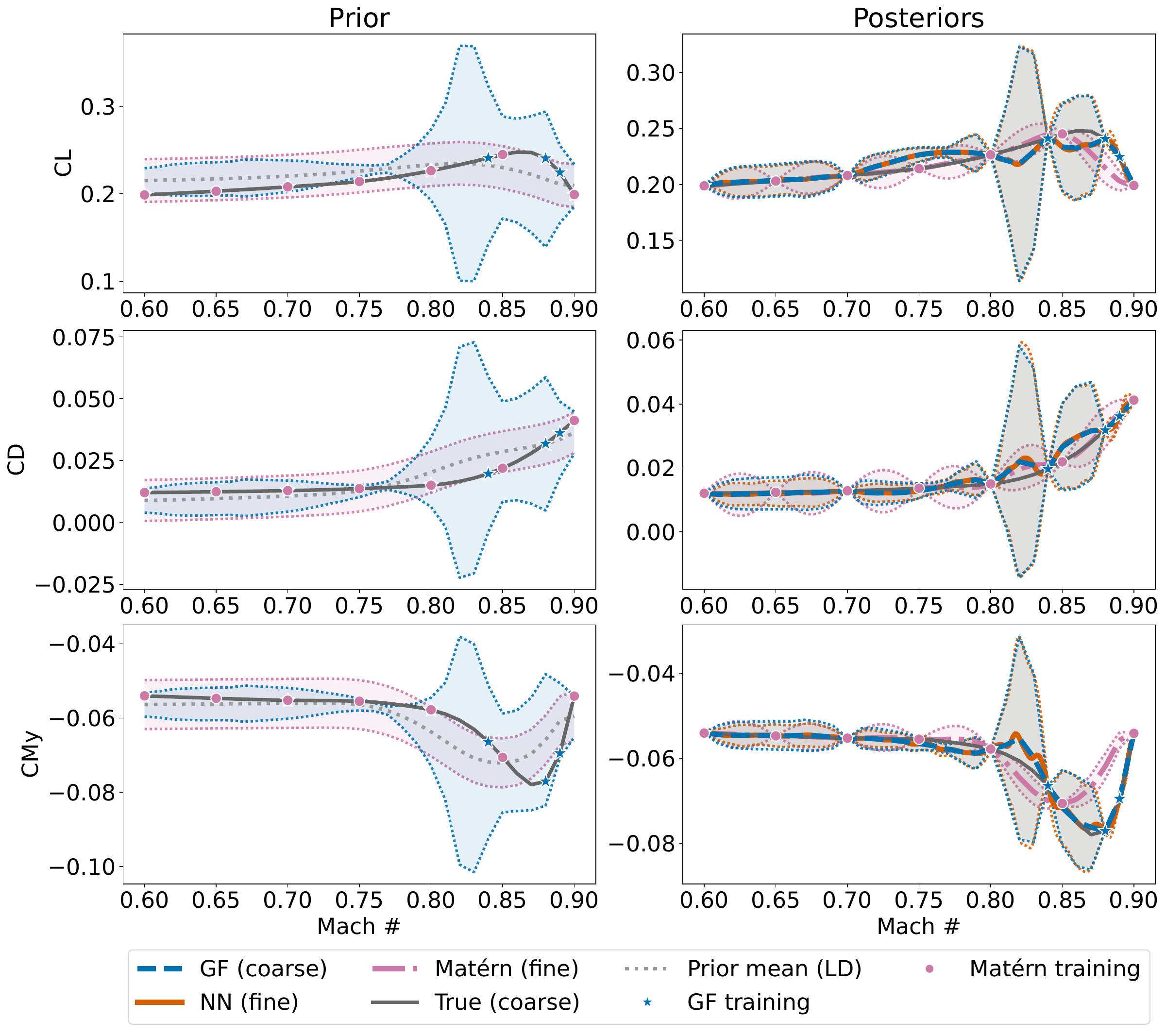}
        
    \caption{Comparison between our GF kernel and Mat\'ern mixture on the QoI regression at iteration 3.}
    \label{fig:mach_adapt_3}
\end{figure*}

\begin{figure*}[htp]
    \centering
    \includegraphics[width=\textwidth]{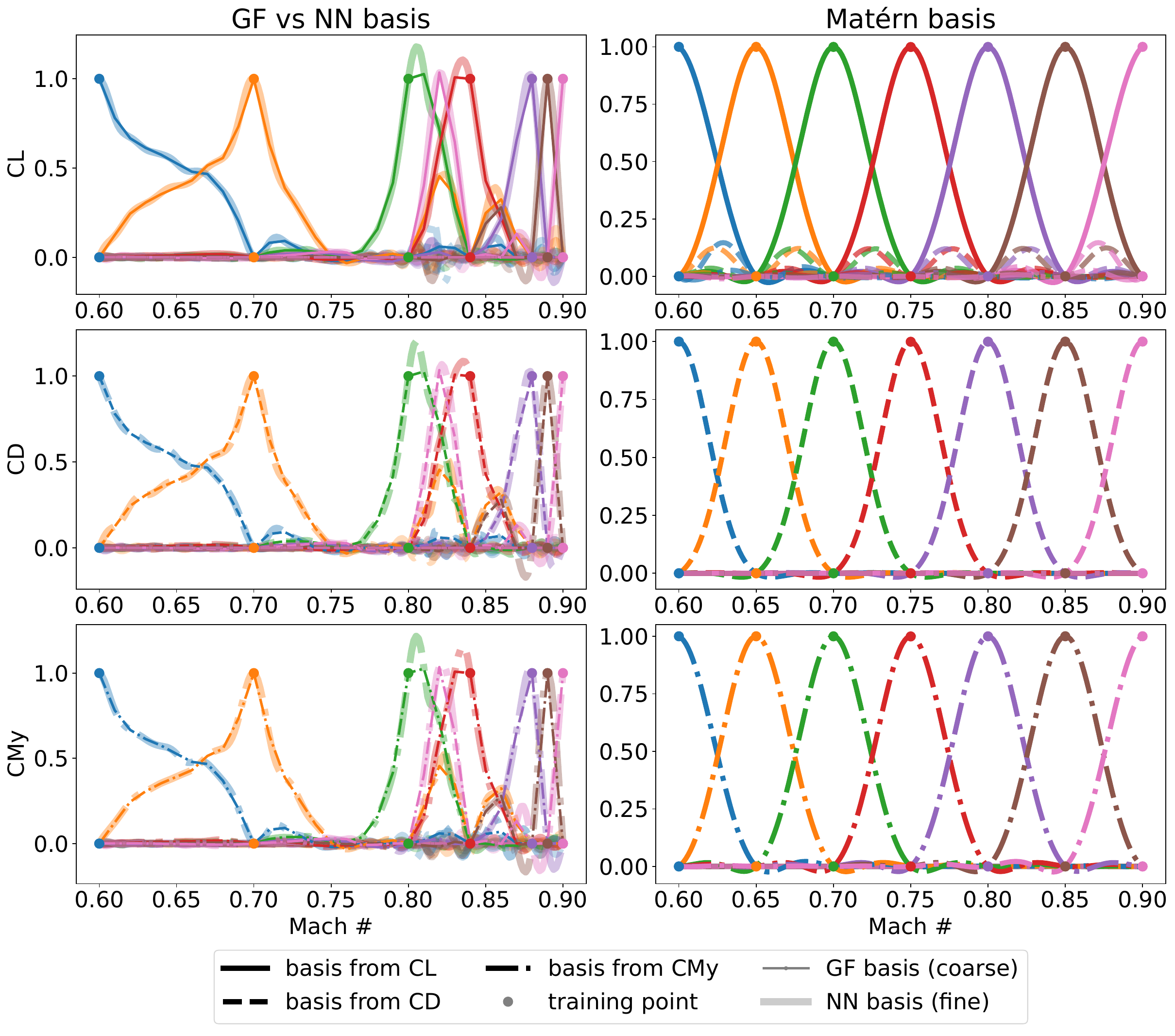}    
    \caption{Comparison between our GF kernel and Mat\'ern mixture basis functions at iteration 3.}
    \label{fig:mach_adapt_3_bases}
\end{figure*}

We repeat the presentation of regressions and associated basis functions for iterations 0 and 3 in Figures~\ref{fig:mach_adapt_0}\textendash\ref{fig:mach_adapt_3_bases}. These also include the neural network kernel, using hyperparameters detailed later.

\subsubsection{Cost-Benefit Analysis}
To quantify the cost-benefit tradeoff of using the PDE kernels to predict LD QoIs compared to generating the HD data, we compare wall clock evaluation time. Across all parameterizations, the HD primal solutions ran for an average of 54,700 s, the LD primal solutions ran for an average of 92 s, and the LD adjoint solutions ran for an average of 162 s. We also observed over 200 parameterizations, the average cost to evaluate the neural network kernel was 0.498 $\mu$s per pair (matrix entry) and 99.6 $\mu$s per parameter. Accounting for the fact that the primal solution can be re-used across QoIs, the NN provides a speedup of $1.93e6$ when evaluating our three QoIs at a new parameter. All cases were executed on a single Intel Xeon Platinum 8362 CPU with 64 processes, fully utilizing its available threads.

As shown in Figures \ref{fig:adapt_0}-\ref{fig:adapt_3}, using the PDE kernel resulted in a better accuracy for a given number of adaptive iterations as shown in the varying winglet size case. This demonstrates the ability to save on HD evaluation costs as fewer samples are required to achieve a given accuracy. In the Mach-varying case, the PDE kernel more closely followed the HD QoI response and gave more informative uncertainty compared to the stationary kernel given samples at identical parameters, further strengthening its benefit in adaptive sampling applications.

While case dependent (more expensive 3D solvers will exaggerate the benefits; poor 2D solvers will diminish the benefits), we
empirically demonstrate the capability of the PDE kernels to give improved regression capabilities at an overall cost savings for expensive
many-query applications.
\subsection{Neural Network Accelerated Kernel}

\begin{figure*}
    \centering
    \begin{tabular}{cc}
        \STAB{\includegraphics[width=1.0\textwidth]{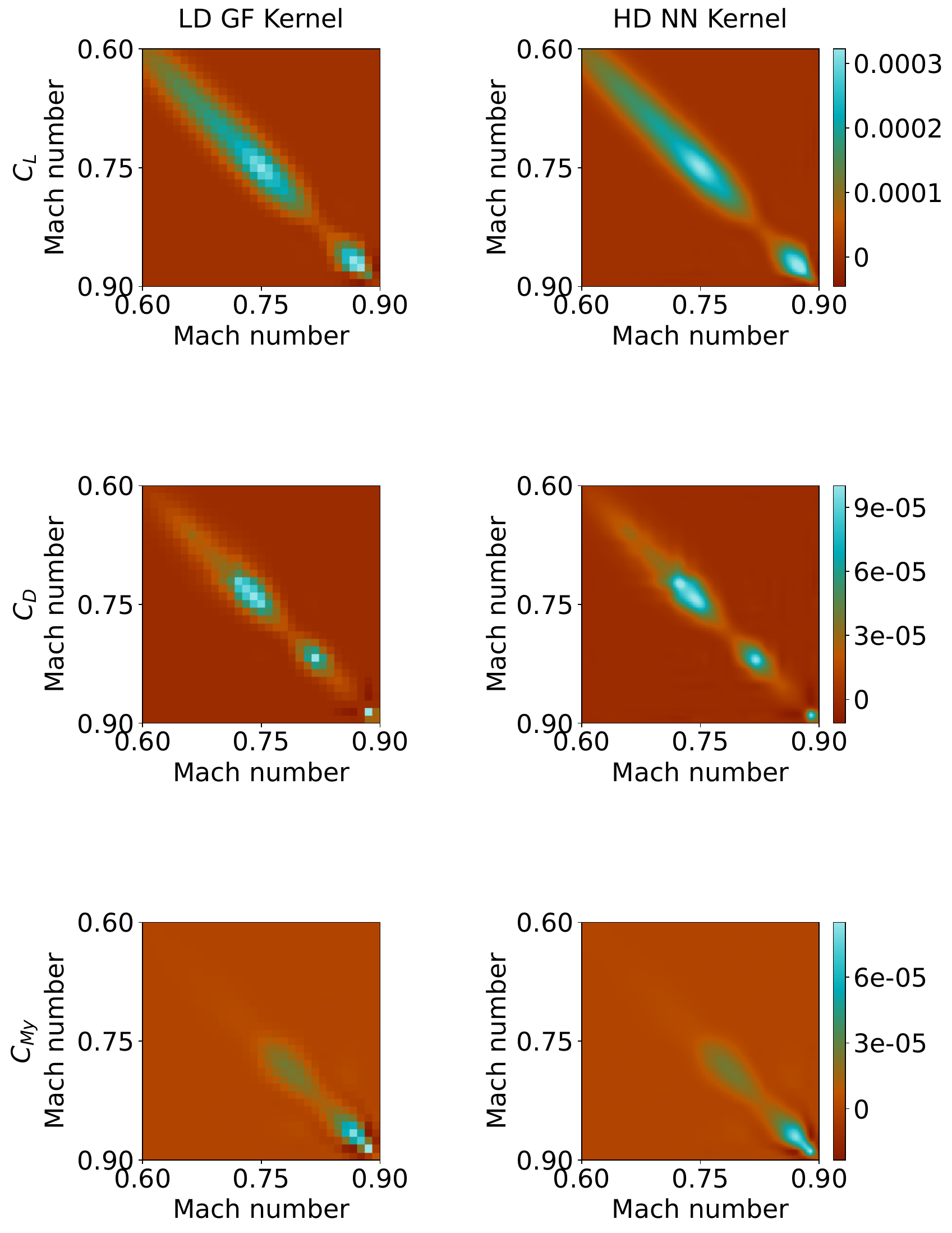}}
    \end{tabular}
    \caption{GF kernel and neural network kernel interpolation}
    \label{fig:k_nn}
\end{figure*}

We use a standard deep, fully-connected neural network. To aide in learning finer scale features, we use a Gaussian Fourier feature mapping, i.e. inputs $\bs{\theta}$ are mapped to $[\cos(2 \pi B \bs{\theta}), \sin(2 \pi B \bs{\theta})]^\T$, where $B\in \mbb{R}^{F \times P}$ is a random matrix with entries $B_{i,j} \sim \mc{N}(0,\sigma^2)$. We chose $F=8,\sigma=1.0$ and note that the parameters (e.g., Mach number) are first scaled to a unit domain. We see, given otherwise identical neural networks trained separately, that final error in the NN results converges quickly with $M$ as detailed in Table~\ref{table:M_convergence}. We emphasize that the displayed relative errors are on the shifted dataset and can be several orders of magnitude smaller on the bare kernel evaluations due to the flatness of the GF kernel functions. The final neural network architecture used 3 hidden layers, each with 512 neurons, 64 series terms, and 8 random Fourier features. The same settings were used for each QoI. The NN basis functions evaluated at significantly more points than the LD model are plotted in transparent curves in under the GF basis functions. The GF kernel and the NN interpolated kernel are compared for the second Mach iteration in Figure~\ref{fig:k_nn}

In addition to accurately replicating the kernel evaluations, as discussed in the Varying Flight Condition results, the NN is able to adequately replicate the GP posterior when used for all involved kernel evaluations.
This requires a particularly small error as small differences in entries of the training data matrix can have a significantly larger effect on the solved coefficients.
Nevertheless, the NN was able to provide sufficient accuracy.
The speedup in the order of millions justifies its use in real-time applications but it may be unnecessary when only a few LD parameters need be evaluated with no demanding time constraints.

\begin{table}[!ht]
\centering
\caption{Post-training error over the entire dataset, as $M$ increases for $C_L$ in the varying-mach case}
\begin{tabular}{lc}
    \hline
    M & Relative Element-Wise L2 Error \\
    \hline
    1 & 9.7E-2 \\
    2 & 4.0E-4 \\
    4 & 2.5E-4 \\
    8 & 2.1E-4 \\
    16 & 1.8E-4 \\
    32 & 1.6E-4 \\
    64 & 1.7E-4 \\
    \hline
\end{tabular}
\label{table:M_convergence}
\end{table}

\section{Conclusion}\label{sec:conclusion}
We introduced a warped inner product kernel over parameter space that uses LD primal and adjoint solutions as feature maps for the parameters. Our novel contributions were the application of GF to nonlinear models and observation functionals, with differing model dimensionality, and our approach for speeding up kernel evaluations with neural networks.
This construction creates a non-stationary kernel that attenuates covariance and increases variance where the LD model has higher sensitivity. The resulting kernel provides a better basis for regressing the response surface and confidence interval with minimal HD data. To account for geometry changes that a primary LD model cannot capture (e.g., symmetry-breaking features), we introduced auxiliary LD models to featurize these parameters. The resulting product kernel effectively captures variation across the entire design space. We proposed a machine learning approach to learn the featurized Mercer kernel. This approach speeds up the kernel evaluation substantially by eliminating the need for forward and adjoint LD solves, which could become expensive during online inference. The non-stationary PDE-informed kernel resolved complex features, such as the rapid change in aerodynamic forcing at the critical Mach number and small winglet height, which stationary kernels failed to capture without further sampling. Ultimately, the integration of dimension-bridging GF corrections with neural network surrogates enables real-time predictions of quantities of interest.

\section*{Appendix}
\section{Approximating Error with Adjoint Inner Products}\label{sec:app_a}
First we show how the adjoint solution accounts for the error in the LD QoI prediction.
To that end, we frame the correction problem \eqref{eq:corrected_ld_pde} as a constrained
least squares problem (the so-called optimal recovery problem in the kernel ridge regression
approach). 

\begin{equation}
    \begin{aligned}
        \underset{g}{\mathrm{minimize}}\ &\underset{=: \mathcal{J}\LRp{\boldsymbol{g},\boldsymbol{u}^\dagger}}{\underbrace{\frac{1}{2}\nor{\boldsymbol{d} - {q}^\dagger\LRp{\boldsymbol{\theta}}}_{\Sigma_\eta^{-1}}^2 + \frac{1}{2}\nor{g\LRp{\boldsymbol{\theta}}-m\LRp{\boldsymbol{\theta}}}_{\mathcal{H}_k}^2}}\\
        \mathrm{s.t.}\ &\begin{cases}
            \underset{=: R\LRp{g,u^\dagger,\boldsymbol{\theta}}}{\underbrace{\mathcal{P}_L\LRp{u^\dagger} - f_L - g\LRp{\boldsymbol{\theta}}}} = 0 \quad \forall \theta \in T \\
            q^\dagger\LRp{\theta} := F\LRp{u^\dagger\LRp{\theta}}
        \end{cases}
    \end{aligned}
    \label{eq:least_sq_constrained}
\end{equation}

where, again, $q^\dagger\LRp{\theta}$ is the QoI given by the corrected solution $u^\dagger$
for parameter $\theta$ and correction $g\LRp{\theta}$, $\Sigma_\eta = \sigma_\eta^2 I$ is the
covariance of the observation noise, $\nor{\cdot}_{\mathcal{H}_k}$ is the norm in the Reproducing
Kernel Hilbert Space (RKHS) generated by the prior GF covariance kernel $k$.

Note that the optimization problem \eqref{eq:least_sq_constrained} could be shown to be a MAP (maximum a posteriori) point of a Bayesian formulation, but we omit the details here. We are not interested in finding the MAP point, but to provide a constructive
approach for deriving the adjoint equation to find the test function $\phi$, which we need for our functional 
Gaussian process regression. To that end we consider the following artificial optimal recovery problem:

\begin{equation}
    \begin{aligned}
        \underset{g}{\mathrm{minimize}}\ \mathcal{J}\LRp{\boldsymbol{g},\boldsymbol{u}^\dagger} &:= F\LRp{u^\dagger\LRp{\boldsymbol{\theta}}}\\
        \mathrm{s.t.}\, R\LRp{g,u^\dagger,\boldsymbol{\theta}} &= 0 \quad \forall \theta \in T 
    \end{aligned}
    \label{eq:least_sq_constrainedA}
\end{equation}



We derive the adjoint equation using the Lagrangian approach. To begin, we 
define an unconstrained
optimal recovery problem with the Lagrangian $\mathcal{L}$.

\begin{equation}
    \underset{g,u^\dagger,\lambda}{\mathrm{minimize}}\ \ \mathcal{L} := \mathcal{J}\LRp{\boldsymbol{g},\boldsymbol{u}^\dagger} - \LRp{R\LRp{\boldsymbol{g},\boldsymbol{u}^\dagger}, \boldsymbol{\lambda}}
    \label{eq:least_sq_unconstrained}
\end{equation}

At any critical point of \eqref{eq:least_sq_unconstrained}, $\boldsymbol{g}$ and $\bs{\lambda}$ must satisfy the following forward and adjoint equation in the weak form:

\begin{equation}
    \begin{gathered}
        \LRp{\mathcal{P}_L\LRp{\boldsymbol{u}^\dagger} - f_L - \boldsymbol{g}, \delta\boldsymbol{\lambda}} = 0 \\
        \LRp{\frac{\delta F}{\delta \boldsymbol{u}^\dagger}, \delta \boldsymbol{u}^\dagger}
        - \LRp{\LRp{\frac{\delta \mathcal{P}_L}{\delta \boldsymbol{u}^\dagger}}^*\!\boldsymbol{\lambda}, \delta \boldsymbol{u}^\dagger} = 0
    \end{gathered}
    \label{eq:first_der_cond}
\end{equation}

where we have defined $\frac{\delta F}{\delta \boldsymbol{u}^\dagger}$ and $\frac{\delta \mathcal{P}_L}{\delta \boldsymbol{u}^\dagger}$ as the functional derivatives of $F$ and $\mathcal{P}_L$ with respect to $\boldsymbol{u}^\dagger$, and $\delta \boldsymbol{u}^\dagger$ as an arbitrary variation of $\boldsymbol{u}^\dagger$. The functional derivatives and variations of other quantities are defined correspondingly in a similar fashion. Here, $\LRp{\frac{\delta \mathcal{P}_L\LRp{\boldsymbol{u}^\dagger}}{\delta \boldsymbol{u}^\dagger}}^*$ denotes the adjoint (with respect to the $L^2$-inner product) of $\frac{\delta \mathcal{P}_L\LRp{\boldsymbol{u}^\dagger}}{\delta \boldsymbol{u}^\dagger}$. 
In the strong form, the forward and the adjoint equations read

\begin{gather*}
       \mathcal{P}_L\LRp{\boldsymbol{u}^\dagger} = f_L + \boldsymbol{g}  \\
         \LRp{\frac{\delta \mathcal{P}_L}{\delta \boldsymbol{u}^\dagger}}^*\!\boldsymbol{\lambda} =  \frac{\delta F}{\delta \boldsymbol{u}^\dagger}
\end{gather*}

We are interested in estimating the error in QoI:

\begin{equation*}    
    q^\dagger\LRp{\bs{\theta}} - q_L\LRp{\boldsymbol{\theta}} = F\LRp{u^\dagger\LRp{\theta}} - F\LRp{u_L\LRp{\theta}} \approx \LRp{\frac{\delta F}{\delta u_L}, \bs{u}^\dagger\LRp{\theta} - \bs{u}_L\LRp{\theta}},
\end{equation*}

where we have ignored $\mc{O}\LRp{\nor{u^\dagger-u_L}^2}$ in the Taylor expansion of $F\LRp{u^\dagger\LRp{\theta}}$ around $u_L$. To the end of this section, we will ignore all the second-order terms, including cross second-order terms such as $\LRp{\frac{\delta F}{\delta \boldsymbol{u}^\dagger} - \frac{\delta F}{\delta \boldsymbol{u}_L}, u^\dagger-u_L}$. We continue estimating the error using the strong form adjoint equation:

\begin{align*}
q^\dagger\LRp{\theta} - q_L\LRp{\theta} &\approx
\LRp{\frac{\delta F}{\delta \bs{u}^\dagger}, \bs{u}^\dagger\LRp{\theta} - \bs{u}_L\LRp{\theta}} \\ &= 
\LRp{\LRp{\frac{\delta \mathcal{P}_L}{\delta \boldsymbol{u}^\dagger}}^*\!\boldsymbol{\lambda}, \bs{u}^\dagger\LRp{\theta} - \bs{u}_L\LRp{\theta}} \\
&\approx \LRp{\LRp{\frac{\delta \mathcal{P}_L}{\delta \boldsymbol{u}_L}}^*\!\boldsymbol{\lambda}, \bs{u}^\dagger\LRp{\theta} - \bs{u}_L\LRp{\theta}}  \\ &= 
\LRp{\boldsymbol{\lambda}, \frac{\delta \mathcal{P}_L}{\delta \boldsymbol{u}_L}\LRp{\bs{u}^\dagger\LRp{\theta} - \bs{u}_L\LRp{\theta}}}
\end{align*}

In the third line, we have linearized about $\bs{u}_L$, and thus assume the functional derivative $\frac{\delta \mathcal{P}_L}{\delta \boldsymbol{u}^\dagger}$ is well-approximated by $\frac{\delta \mathcal{P}_L}{\delta \boldsymbol{u}_L}$. This is a strong assumption that emphasizes the importance of selecting an LD model which balances cost and accuracy. Now from the strong form of the forward equation, similarly linearizing about $\bs{u}_L$ gives

\[
\mathcal{P}_L\LRp{\boldsymbol{u}_L} + \frac{\delta \mathcal{P}_L}{\delta \boldsymbol{u}_L}\LRp{\bs{u}^\dagger\LRp{\theta} - \bs{u}_L\LRp{\theta}} \approx f_L + \boldsymbol{g}
\]

and since $\mathcal{P}_L\LRp{\boldsymbol{u}_L} = f_L$, we have

\[
\frac{\delta \mathcal{P}_L}{\delta \boldsymbol{u}_L}\LRp{\bs{u}^\dagger\LRp{\theta} - \bs{u}_L\LRp{\theta}} \approx \boldsymbol{g}
\]

The error estimation for the QoI now reads

\begin{equation}
    q^\dagger\LRp{\bs{\theta}} - q_L\LRp{\boldsymbol{\theta}} \approx
    \LRp{\bs{\lambda},\bs{g}}
    \label{eq:adj_inpr}
\end{equation}

When the forward equation is linear, i.e. when $\mathcal{P}_L$ is a linear operator, and  the quantity of interest is linear, i.e when $F$ is a linear operator, then there are no approximation errors in the above derivation. In particular, we would have

\[
q^\dagger\LRp{\bs{\theta}} - q_L\LRp{\boldsymbol{\theta}} =
\LRp{\bs{\lambda},\bs{g}}
\]

\subsection{Regression}
We need to compute the inner products \eqref{eq:adj_inpr} for a given $\bs{\theta}$ for which we do not have HD observations (i.e., we do not know $q^\dagger$, and thus cannot find $\LRp{\bs{\lambda},\bs{g}}$).
We assume that the noise and the GF $\bs{g}$ are independent. The problem of finding the
posterior distribution of $\bs{g}$ given the noisy error observation $q^\dagger\LRp{\bs{\theta}} - q_L\LRp{\bs{\theta}}$, where $q^\dagger\LRp{\bs{\theta}}$ on the training data set is given as in \eqref{eq:noisy_match}, can be
derived analogously to the standard Gaussian process regression since

\[
d - q_L\LRp{\bs{\theta}} = \LRp{\bs{\lambda},\bs{g}} + \eta
\]

 In particular, we can define a 
multivariate Gaussian distribution describing the corrective GF $g$ by its evaluation on a finite set
of adjoint-parameter pairs. We then define a joint distribution between the evaluations on the test
set and train set 
to give the covariance between the observed error on a point in
the training set $d\LRp{\theta} - q_L\LRp{u_L\LRp{\theta}}$ and a prediction on the test set 
$\LRp{\phi\LRp{\hat{\theta}},g\LRp{\hat{\theta}}}$ because they are modeled as jointly Gaussian.

The posterior (or conditional) GF has a posterior mean $m^\dagger$ and covariance $k^\dagger$ given 
by formulas analogous to the multivariate conditional distribution. 

\begin{equation}
    \begin{gathered}
        g|\boldsymbol{d} \sim \mathcal{GF}\LRp{m^\dagger,k^\dagger}\\
        m^\dagger = m + \boldsymbol{k}_{T}\LRp{K_{TT}+\Sigma_\eta}^{-1}\LRp{\boldsymbol{d} - \boldsymbol{q}_L}\\
        k^\dagger = k - \boldsymbol{k}_{T}\LRp{K_{TT}+\Sigma_\eta}^{-1}\boldsymbol{k}_{T}^{\T}\\
        \boldsymbol{k}_{T} = \begin{bmatrix}
            k\LRp{\cdot,\LRp{\phi\LRp{\theta_1},\theta_1}}\\
            \vdots\\
            k\LRp{\cdot,\LRp{\phi\LRp{\theta_{N_T}},\theta_{N_T}}}\\
        \end{bmatrix}
    \end{gathered}
    \label{eq:GF_post}
\end{equation}

where the capital $K$ denotes the covariance matrix generated from evaluating the covariance operator
on the test functions (and associated parameters) in the subscript and $\boldsymbol{k}_{T}$ is the
vector of kernel functions associated with the train tuples. We then use the posterior to
predict new QoIs at new parameter configurations by running the adjoint solver at the new configuration $\hat{\theta}$
(if not run already) and evaluate 

\begin{equation}
    q^\dagger\LRp{\hat{\theta}} = q_L\LRp{\hat{\theta}} + m^\dagger\LRp{\phi\LRp{\hat{\theta}},\hat{\theta}}
\end{equation}
\label{app:adj_map}
\section{Derivation of the Adjoint Inner Product Kernel}
Using the standard definition of covariance and the postulation $g \sim \mathcal{GF}\LRp{m,k_V\!}$, we write the covariance between the evaluation of the GF on two functions as 

\begin{equation}
    \begin{aligned}
         k_V\LRp{v,w} &= \mathrm{Cov}\LRs{g\LRp{v},g\LRp{w}} \\
            &= \mbb{E}\LRs{\LRp{g\LRp{v}-m\LRp{v}}\LRp{g\LRp{w}-m\LRp{w}}} \\
            &= \mbb{E}\LRs{\LRp{g-m,v}\LRp{g-m,w}}
    \end{aligned}
    \label{eq:cov_expectation}
\end{equation}

Here, we use the blackboard $\mbb{E}$ to denote the expectation operator. In \eqref{eq:cov_expectation}, we identify the
functionals with their function representers in the appropriate function space and combine the inner products with the linearity
of the inner product. Note that taking the test functions $v,w$ to be the adjoint states $\lambda$ and $\lambda'$ from Appendix \ref{sec:app_a}, we see that the kernel predicts the covariance in the errors of the LD model.

\begin{equation}
    k_V\LRp{\lambda,\lambda'} \approx \mathrm{Cov}\LRs{\LRp{q^\dagger-q_L},\LRp{q^{\dagger}-q_L}'}
\end{equation}

Rewriting the $L^2$ inner product with integrals over the appropriate domain and combining them with Fubini's Theorem, we now have
the expression 

\begin{equation}
k_V\LRp{v,w} \\=       \iint \underset{=: \kappa\LRp{x,y}}{\underbrace{\mbb{E}\LRs{\LRp{g\LRp{x}\!-\!m\LRp{x}}\LRp{g\LRp{y}\!-\!m\LRp{y}}}}}\ v\LRp{x}\!dx\, w\LRp{y}\!dy
\label{eq:gf_cov_inner}
\end{equation}

The expression for $k_V\LRp{v,w}$ in \eqref{eq:gf_cov_inner} also tells us that we can define $k_V\LRp{v,w}$ by first choosing an appropriate kernel $\kappa\LRp{x,y}$. Indeed, if $\kappa$ is a positive definite Hilbert-Schmidt kernel, from Mercer's Theorem \cite{mercer1909xvi}, we know that continuous and bounded positive definite kernels uniquely 
identify Hilbert-Schmidt integral operators. Thus, $k_V\LRp{v,w}$ defined through a Hilbert-Schmidt kernel $\kappa\LRp{x,y}$ via the integral \eqref{eq:gf_cov_inner} is indeed symmetric positive definite. This approach allows us to use kernels of GPs to define kernels for GFs, which simplifies the construction of GF kernels significantly.

For vector-valued function spaces, we rewrite the integral (assuming column vectors) in \eqref{eq:gf_cov_inner} as

\begin{equation}
    \iint \bs{v}\LRp{x}^{\T}\!K\LRp{x,y}\bs{w}\LRp{y} dxdy
    \label{eq:vecval_gf_cov_inner}
\end{equation}

where the integral kernel is now denoted with $K$, a symmetric positive definite matrix. This form was similarly justified in \cite{sung2024}. 

A general nonstationary parameterization of $K$ based on LU factorization can be given as follows:

\begin{subequations}
    \begin{gather}
        K\LRp{x,y} = L\LRp{x}L\LRp{y}^{\T} \\
        L\LRp{x}_{ij} = \begin{cases}
            0 & i < j \\
            \varphi_{ij}\LRp{x} & i \geq j
        \end{cases}
    \end{gather}
\end{subequations}

where each parametric element function $\varphi_{ij}$ is continuous and bounded. This allows for asymmetric cross covariance
matrices, but ensures collocated covariance matrices are positive semidefinite. If we use a stationary kernel, we would no longer
be able to differentiate between transposed cross evaluations, so cross covariances must also be symmetric.

Note that stationary GPs do not represent stationary GFs in general. This is because a stationary GF requires its covariance
operator to be a function of the distance in function space $d_V\!=\!\nor{v-w}_{L^2}$, whereas the stationary GP has a covariance that is
a function of distance in the physical domain $d_\Omega\!=\!\nor{x-y}_{\ell^2}$. 
Consider a set of test functions with compact support over
a small subset of the domain that have translational symmetry. For any two functions in this set, if their supporting sets
are disjoint almost everywhere, their distance in the function space is the same as all other such pairs.
Thus, a stationary GF kernel would have equal covariance between evaluations on all such function pairs. However, a GF represented by a
stationary GP will predict smaller covariance as the translation distance grows.

For this reason, we opt to use the stationary formulation for the GP covariance. We use the Linear Model for Coregionalization (LMC)
from geostatistics \cite{bourgault1991multivariable}. LMC defines $K$ as a (physically) stationary diagonally weighted product of two
full column rank matrices.

\begin{subequations}
    \begin{gather}
        K\LRp{x,y} = AD\LRp{x,y}\!A^{\T} \\
        D_{ii}\LRp{x,y} = s_i\LRp{\nor{x-y}} \\
        A \in \mbb{R}^{n \times m} ,\ n \geq m
    \end{gather}
\end{subequations}

where $s_i$ are possibly independently parameterized stationary kernels that define the diagonal
elements for any pair of evaluation points.

For parameter efficiency, we define $A$ as a lower triangular matrix, which uses half of the parameters to factor a full rank $K$.
This is equivalent to defining a stationary Cholesky factorization by absorbing $D^{1/2}$ into $L$.

\begin{subequations}
    \begin{gather}
        K\LRp{x,y} = LD\LRp{x,y}\!L^{\T} = \tilde{L}\LRp{x,y}\tilde{L}\LRp{x,y}^{\T} \\
        L_{ij} = \begin{cases}
            0 & i < j \\
            1 & i = j \\
            \lambda_{ij} & i > j
        \end{cases} \\
        \lambda_{ij} \in \mbb{R}
    \end{gather}
\end{subequations}\label{app:kernel}

\section*{Funding Sources}
This manuscript has been authored, in part, by UT-Battelle, LLC under Contract No.~DE-AC05-00OR22725 with the U.S.~Department of Energy.
The United States Government retains and the publisher, by accepting the article for publication, acknowledges that the United States Government retains a non-exclusive, paid-up, irrevocable, world-wide license to publish or reproduce the published form of this manuscript, or allow others to do so, for United States Government purposes. The Department of Energy will provide public access to these results of federally sponsored research in accordance with the DOE Public Access Plan (\url{http://energy.gov/downloads/doe-public-access-plan}).

\section*{Acknowledgment}
We acknowledge the Texas Advanced Computing Center (TACC) at The University of Texas at Austin for providing computational resources that have contributed to the research results reported within this paper. URL: \href{http://www.tacc.utexas.edu }{http://www.tacc.utexas.edu }.


\bibliography{refs}


\end{document}